\documentclass[preprint,12pt]{article}
\usepackage{rotating}
\usepackage{epsfig}
\usepackage{ifpdf}
\usepackage{dsfont}
\usepackage{amssymb}
\usepackage{lscape}

\usepackage{amssymb}

\newtheorem{theorem}{Theorem} 
\newtheorem{corollary}{Corollary}

\newtheorem{lemma}{Lemma}
\newtheorem{proposition}{Proposition}
\newtheorem{rem}{\bf Remark}

\def\1{1\kern-.20em {\rm l}}

\newcommand{\R}{\mathbb{R}}
\newcommand{\T}{\mathcal{T}}
\newcommand{\G}{\mathcal{G}}
\newcommand{\F}{\mathcal{F}}
\newcommand{\N}{\mathcal{N}}
\newcommand{\E}{\mathbb{E}}
\newcommand{\C}{\mathcal{C}}

\title{{\bf Rate of uniform consistency for a class of
mode regression on functional stationary  ergodic data. Application to electricity consumption}}  %

\author{\bf  Mohamed Chaouch$^a$,    Na\^amane La\"ib$^b$  \&  Djamal Louani$^{b,*}$
        \\ 
{\small  $^a$ Department of Statistics, UAE University, United Arab Emirates}
\\
 {\small $^b$ L.S.T.A.,
Universit\'e de Paris 6, France.}\\
 }

\date{}
\begin{document}
\maketitle
\begin{quote}
\noindent {\bf Abstract} - The aim of this paper is to study  the
asymptotic properties of a class of kernel conditional mode
estimates whenever functional stationary ergodic data are
considered. To be more precise on the matter, in the ergodic data
setting, we consider a random element $(X, Z)$ taking values in
some semi-metric abstract space $E\times F$. For a real function
$\varphi$ defined on the space $F$ and $x\in E$, we consider the
conditional mode of the real random variable $\varphi(Z)$ given
the event $``X=x"$. While estimating the conditional mode
function, say $\theta_\varphi(x)$,  using the well-known kernel
estimator, we establish the  strong consistency with rate of this
estimate uniformly over Vapnik-Chervonenkis classes of functions
$\varphi$. Notice that the ergodic setting offers a more general
framework than the usual mixing structure. Two applications to energy data are provided to illustrate some
examples of the proposed approach in time series forecasting
framework. The first one consists in forecasting the {\it daily
peak} of electricity demand in France (measured in Giga-Watt). Whereas the second one
deals with the short-term forecasting of the electrical {\it energy} (measured in Giga-Watt per Hour)
that may be consumed over some  time intervals that cover the peak
demand.


\vspace{1mm}

\vspace{1mm}

\end{quote}

\clearpage
\setcounter{section}{1}
\centerline{\large 1. INTRODUCTION}
\vspace{2mm}

\noindent Let $(X,Z)$ be a $E\times F$-valued random elements,
where $E$ and $F$ are some semi-metric abstract spaces. Denote by
$d_E$ and $d_F$ semi-metrics associated to spaces $E$ and $F$
respectively. Let ${\cal C}$ be a class of real functions defined
upon $F$. Obviously, for any $\varphi\in{\C}$, $\varphi(Z)$ is a
real random variable. Suppose now that we observe a sequence
$(X_i,Z_i)_{i\geq 1}$ of copies of $(X,Z)$ that we assume to be
stationary and ergodic. For any $x\in E$ and any $\varphi\in{
\C}$, let $g_{\varphi}(.|x)$ be the conditional density of
$\varphi(Z)$ given $X=x$. We assume that $g_{\varphi}(.|x)$ is
unimodal  on some compact $S_\varphi\subset\mathbb{R}$. The
conditional mode is defined, for any fixed $x\in E$, by
$$\Theta_{\varphi}(x)=\mbox{arg}\sup_{y \in S_\varphi} g_{\varphi}(y|x).$$
Note that, if there exists $\xi>0$ such that for any
$\varphi\in\C$
\begin{eqnarray}
g_{\varphi}(.|x)\uparrow\  \mbox{on}\ (\Theta_{\varphi}(x)-\xi,\
\Theta_{\varphi}(x))\  \mbox{and}\ g_{\varphi}(.|x)\downarrow\
\mbox{on}\ (\Theta_{\varphi}(x),\ \Theta_{\varphi}(x)+\xi),
\end{eqnarray}
and if we choose $ S_\varphi= \lbrack\Theta_{\varphi}(x)-\xi,\
\Theta_{\varphi}(x)+\xi\rbrack$, then the mode
$\Theta_{\varphi}(x)$ is uniquely defined for any $\varphi$.
 The kernel estimator,
say $\hat{\Theta}_{\varphi,n}(x)$, of $\Theta_{\varphi}(x)$ may be
defined as the value maximizing the kernel estimator
$g_{\varphi,n}(y|x)$ of $g_{\varphi}(y|x)$, that is,
\begin{eqnarray}\label{def_est}
g_{\varphi,n}(\hat{\Theta}_{\varphi,n}(x)|x)=\sup_{y\in S_\varphi}g_{\varphi,n}(y|x).
\end{eqnarray}
Here,
$$g_{\varphi,n}(y|x)=\frac{{f}_{\varphi,n}(x,y)}{l_n(x)},$$
where
$$f_{\varphi,n}(x,y)\!=\!\!\frac{1}{nh\E\lbrack
\Delta_1(x)\rbrack}\!\!\sum_{i=1}^{n}\!\left\lbrack\!
\Delta_i(x)H\!\left(\!\frac{y-\varphi(Z_i)}{h}\!\right)\!\right\rbrack\!,\!
\ l_n(x)\!=\!\!\frac{1}{n\E\lbrack
\Delta_1(x)\rbrack}\!\sum_{i=1}^{n} \!\Delta_i(x)$$ and
$\displaystyle{\Delta_i(x)=K\left(\frac{d(x,X_i)}{h}\right)}$,
with $K$ and $H$ two real valued kernels and $h:=h_n$ a sequence
of positive real numbers tending to zero as $n\rightarrow\infty$.

The aim of this paper is to establish the uniform consistency,
with respect to the function parameter $\varphi\in{\C}$,
%
of the conditional mode estimator
$\hat{\Theta}_{\varphi,n}(x)$ when data are assumed to be sampled
from a stationary and ergodic process. More precisely, under
suitable conditions upon the entropy of the class ${\C}$ and the
rate of convergence of the smoothing parameter $h$ together with
some regularity conditions on the distribution of the random
element $(X,Z)$, we obtain results of type
$$\sup_{\varphi\in{\C}}|\hat{\Theta}_{\varphi,n}(x)-\Theta_{\varphi}(x)|=O(\alpha_n), \ \mbox{a.s.}$$
where $\alpha_n$ is a  quantity  to be specified later on.
%
 Notice that, besides the infinite dimensional character of the data, the
ergodic framework avoid the widely used strong mixing condition
and its variants to measure the dependency and the very involved
probabilistic calculations that it implies (see, for instance,
Masry (2005)). Further motivations to consider ergodic data are
discussed in La\"ib (2005) and La\"ib \& Louani (2010) where
details defining the ergodic property of processes together with
examples of such processes are also given.

Indexing by a function $\varphi$ allows to consider simultaneously
various situations related to model fitting and time series
forecasting. Whenever $Z:=\{Z(t) : t\in T\}$ denotes a process
defined on some real set $T$, one may consider the following
functionals $\varphi_1(Z)=\sup_{t\in T}Z(t)$ and
$\varphi_2(Z)=\inf_{t\in T}Z(t)$ giving extremes of the process
$Z$ that are of interest in various domains as, for example, the
finance, hydraulics and the weather forecasting. For some weight
function $W$ defined on $T$ and some $p>0$, one may consider the
functional $\varphi_{p,W}$ defined by
$\varphi_{p,W}(Z)=\int_TW(t)Z^p(t)dt$. Further situation is to
consider, for some subset $A$ of $T$, the functional $Z\rightarrow
\varphi_\rho(Z) =\inf\{t\in A : Z(t)\geq \rho\}$ for some
threshold $\rho$. Such a case is very useful in threshold and
barrier crossing problems encountered in various domains as
finance, physical chemistry and hydraulics. Moreover, indexing by
a class of functions $\C$ is a step towards
 modelling a functional response random variable. Indeed, the quantity
 $Z(\varphi):=\{\varphi(Z) : \varphi\in \C\}$ may be viewed as a functional random variable
 offering, in this respect,
 a device for such investigations.
 \vspace{3mm}


\vspace{3mm}

The modelization of the functional variable is becoming more and
more popular since the publication of the monograph of Ramsay and
Silverman (1997) on functional data analysis.  Note however that
the first results dealing with nonparametric models (mainly the
regression function) were obtained  by Ferraty and Vieu (2000).
Since then, an increasing number of papers on this topic  has been
published.
%
%
One may refer to the monograph by Ferraty and Vieu (2006) for an
overview on the subject and the references therein. Extensions to
other regression issues as the time series prediction have been
carried out in a number of publications, see for instance Delsol
(2009). The general framework of ergodic functional data has been
considered by La\"ib and Louani (2010,2011) who stated
consistencies with rates together with the asymptotic normality of
the regression function estimate.\vspace{3mm}

Asymptotic properties of the conditional mode estimator have been
investigated in various situations throughout the literature.
Ferraty {\it et  al.} (2006) studied asymptotic properties of
kernel-type estimators of some characteristics of the conditional
cumulative distribution with particular applications to the
conditional mode and conditional quantiles.  Ezzahrioui and Ould
Sa\"id (2008 and 2010) established the asymptotic normality of the
kernel conditional mode estimator in both i.i.d. and strong mixing
cases. Dabo-Niang and Laksaci (2007) provided a convergence rate
in $L^p$ norm sense of the kernel conditional mode estimator
whenever  functional $\alpha$-mixing observations are considered.
Demongeot {\it et al.} (2010) have established the pointwise and
uniform almost complete convergences with rates of the local
linear estimator of the conditional density. They used their
results to deduce some asymptotic properties of the local linear
estimator of the conditional mode. Attaoui {\it et al.} (2011)
have established the pointwise and uniform almost complete
convergence, with rates, of the kernel estimate of  the
conditional density when the observations are linked with a
single-index structure. They applied their results  to the
prediction problem via the conditional mode estimate. Notice also
that, considering a scalar response variable $Y$ with a covariate
$X$ taking values in a semi-metric space, Ferraty {\it et al.}
(2010) studied, in the i.i.d. case, the nonparametric estimation
of some functionals of the conditional distribution including the
regression function, the conditional cumulative distribution, the
conditional density together with the conditional mode. They
established the uniform almost complete convergence, with rates,
of kernel estimators of these quantities.\vspace{3mm}

It is well-known that the conditional mode provides an alternative
prediction method to the classical approach based on the usual
regression function. Since there exist cases where the conditional
density is such that the regression function vanishes everywhere,
then it makes no sense to use this approach in problems involving
prediction. An example in a finite-dimensional space is given in
Ould-Sa\"id (1997) to illustrate this situation. Moreover, a
simulation study in infinite-dimensional spaces carried out by
Ferraty et al. (2006), shows that the conditional mode approach
gives slightly better results than the usual regression
approach.\vspace{3mm}

\begin{figure}[ht!]
\begin{center}
\includegraphics[height=8cm,width=17cm]{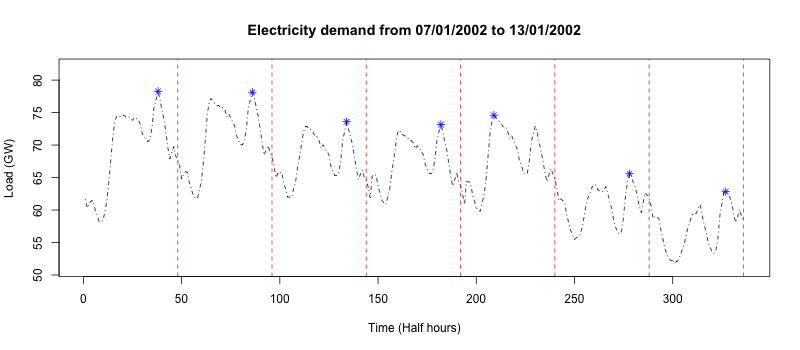}
\end{center}
\caption{Half hour electricity consumption in France from 07/01/2002 to 13/01/2002 (7 days).
The vertical dotted lines separate days and the star points correspond to the
peak demand for each day (in Giga-Watt). }
\label{peak}
\end{figure}

In this paper, two applications to energy data are provided to illustrate
some examples of the proposed approach in time series forecasting framework.
The first real case consists in forecasting the {\it daily peak} of electricity demand in France (measured in Giga-Watt).
Let us denote by $(Z_i(t))_{t\in [0,T]}$ the curve of electricity demand (called also {\it load curve}) measured over un interval [0,T].
If we have hourly (reps. half-hour) measures then $T=24$ (resp. $T=48$).
The peak demand observed for any day $i$ is defined as $\mathcal{P}_i = \sup_{t\in [0,T]} Z_i(t)$.
In such case $\varphi(\cdot)$ is fixed to be the supremum function, over $[0,T]$,
of the function $Z(t)$. Accurate prediction of daily peak load demand is very important for decision in the energy sector.
In fact, short-term load forecasts enable effective load shifting between transmission substations,
scheduling of startup times of peak stations, load flow analysis and power system security studies.
 Figure \ref{peak} provides a sample of seven daily load curves (from 07/01/2002 to 13/01/2002).
 Vertical dotted lines separate days and the star points correspond to the peak demand for each day.

\begin{figure}[ht!]
\begin{center}
\includegraphics[height=10cm,width=17cm]{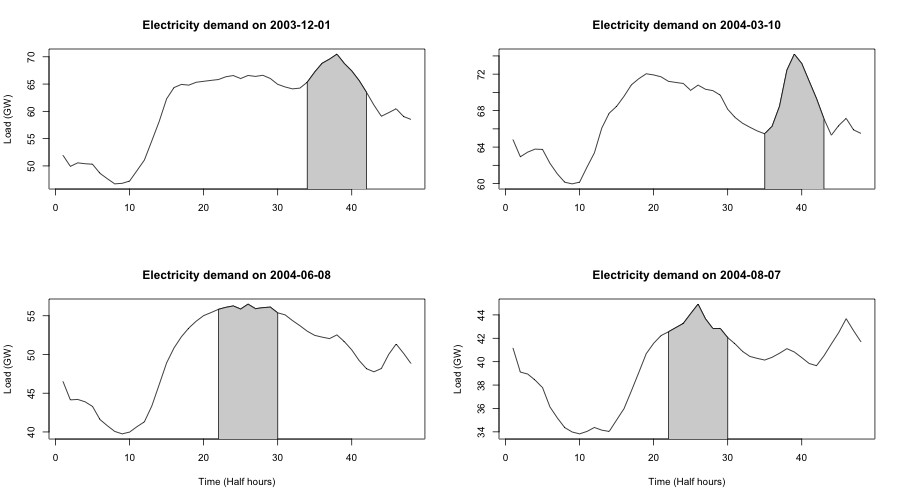}
\end{center}
\caption{A sample of four half hour daily load curves that cover winter and summer seasons.
Solid lines are the daily load curves and the grey surfaces correspond to the electrical energy (in Giga-Watt per Hour)
consumed over an interval of three hours around the peak.}
\label{surface}
\end{figure}

It is well-known that, in addition to peak demand, some other characteristics of the load curve may be of interest
from an operational point of view. In fact the prediction of the {\it electrical energy} (measured in Giga-Watt per Hour)
consumed over an interval of three hours around the peak demand may helps in the determination of consistent
and reliable supply schedules during peak period. Therefore, the second application in this paper deals with the short-term forecasting
of the electrical energy  that may be consumed between  6pm and 9pm in winter and between 12am and 3pm in summer.
Those time intervals cover the peak demand which happens around 7pm in winter and 2pm in summer.
Formally, if we consider $Z_i(t)$ the load curve of some day $i$, then the electrical energy consumed between $t_1$ and $t_2$
is defined as $\mathcal{E}_i = \int_{t_1}^{t_2} Z_i(t) dt$. Therefore, $\varphi(\cdot)$ in that case is the integral function.
A sample of four half hour daily load curves that cover winter and summer seasons is plotted in Figure \ref{surface}.
Solid lines are the daily load curves and the grey surfaces correspond to the electrical energy consumed over an interval of three hours around the peak.


\vspace{5mm}

\setcounter{section}{2}
\centerline{\large 2. RESULTS}
\vspace{2mm}

In order to state our results, we introduce some notations. Let
${\mathcal F}_i$ be  the  $\sigma$-field generated by $(
(X_1,Z_1), \ldots, (X_i, Z_i))$ and  ${\mathcal G}_i$  the one
generated by  $( (X_1,Z_1), \ldots,$ $(X_i, Z_i), X_{i+1})$. Let
$B(x, u)$ be a ball centered at $x\in E$ with radius $u$. Let
$D_i(x):=d(x, X_i)$ so that $D_i(x)$ is a nonnegative real-valued
random variable.  Working on the probability space $(\Omega, {\cal
A}, \mathbb{P})$, let  $F_x(u)=\mathbb{P}(D_i(x) \leq u)
:=\mathbb{P}(X_i\in B(x, u))$ and $F_x^{{\cal
F}_{i-1}}(u)=\mathbb{P}(D_i(x)\leq u \ | {\cal
F}_{i-1})=\mathbb{P}(X_i\in B(x, u)\; | {\cal F}_{i-1})$ be the
distribution function and the  conditional distribution function,
given the $\sigma$-field ${\cal F}_{i-1}$, of $(D_i(x))_{i\geq 1}$
respectively. Denote by $o_{\mbox{a.s.}}(u)$ a real random
function $l$ such that $l(u)/u$ converges to zero almost surely as
$u\rightarrow0$. Similarly, define $O_{\mbox{a.s.}}(u)$ as a real
random function $l$ such that $l(u)/u$ is almost surely bounded.

Our results are stated under some assumptions we gather hereafter
for easy reference.
\begin{enumerate}

    \item [\bf{A1}] For $x \in E$, there exist a sequence of nonnegative random functional $(f_{i,1})_{i\geq
    1}$ almost surely bounded by a sequence of deterministic quantities $(b_i(x))_{i\geq 1}$ accordingly, a sequence
of random functions $(\psi_{i,x})_{i\geq 1}$, a deterministic
nonnegative bounded functional $f_1$ and a nonnegative
nondecreasing real function $\phi$ tending to zero as its argument
goes to zero, such that
\begin{itemize}
        \item[(i)] $\displaystyle{F_x(u)=\phi(u)
        f_1(x)+o(\phi(u))}$, as $u\rightarrow0$,
        \item [(ii)] For any $i\!\in\!\mathbb{N}$,
        $\displaystyle{F_x^{\F_{i-1}}(u)\!=\!\phi(u) f_{i,1}(x)+\psi_{i,x}}(u)$ with $\psi_{i,x}(u)\!=\!o_{a.s.}(\phi(u))$ as
        $u\!\rightarrow\!0$, $\displaystyle{\frac{\psi_{i,x}(u)}{\phi(u)}}$ is
        almost
        surely bounded and
        $\displaystyle{\frac{1}{n}\!\sum_{i=1}^{n}\!\psi_{i,x}(u)\!=\!o_{a.s.}(\phi(u))}$
        as $n\rightarrow\infty$, $u\rightarrow 0$.
        \item [(iii)]$\displaystyle{n^{-1}\sum_{i=1}^{n}f_{i,1}(x)}\rightarrow
        f_1(x)$ almost surely as
        $n\rightarrow \infty$.
        \item [(iv)]There exists a nondecreasing bounded function
        $\tau_0$ such that, uniformly in $u\in\lbrack 0,\
        1\rbrack$,\\
        $\displaystyle{\frac{\phi(hu)}{\phi(h)}\!=\!\tau_0(u)\!+\!o(1)}$, as $h\downarrow0$
        and, for $1\!\leq\!j\!\leq 2$, $\displaystyle{\int_{0}^{1}\!\!(K^j(u))^{\prime}\tau_0(u)du\!<\!\infty}$.
        \item [(v)]  $n^{-1}\sum^n_{i=1}b_i(x)\rightarrow D(x) < \infty$ as $n\rightarrow
        \infty$.\\
    \end{itemize}

    \item[\bf{A2}]  $K$ is a nonnegative bounded kernel of class $\C^1\!$ over its support $\lbrack 0,1\rbrack$,  with $K(1)\!>\!0$ and
     the derivative $K^{\prime}\!$ is such that $K^{\prime}\!(t)\!<\!0$, for any
    $t\in\lbrack 0,1\rbrack$.  \\

    \item [\bf{A3}\ ]
    \begin{itemize}
\item [(i)]For any $\epsilon > 0$, there exists $\eta> 0$ such
that for any $(\varphi_1,\ \varphi_2)\in\C\times\C$,
$d_{\C}(\varphi_1,\ \varphi_2)<\eta$
 implies
that $|\Theta_{\varphi_1}(x)- \Theta_{\varphi_2}(x)|< \epsilon$.

        \item [(ii)] Uniformly in $\varphi\in \C$, $g_\varphi(.|x)$ is uniformly continuous on $S_\varphi$.

        \item [(iii)] $g_{\varphi}(.|x)$ is
 differentiable up to order $2$ and
 $\lim_{n\to \infty}\displaystyle{\sup_{\varphi \in \C }|g_\varphi^{(2)}(\hat{\Theta}_{\varphi,n}(x)|x)|\!:=\!\Phi(x)\!\neq \!0}$.
   \item [(iv)] For any $x\in E$, there exist $V(x)$ a neighborhood of $x$,  some constants $C_x>0$, $\beta>0$ and
    $\nu\in (0,\ 1\rbrack$, independent of $\varphi$, such that for any $\varphi\in\C$, we have
    $\forall\ (y_1,\ y_2) \in S_\varphi\times  S_\varphi$, $\forall (x_1,\ x_2) \in V(x)\times
    V(x)$,\\ \vspace{-1mm}

$|g^{(j)}_{\varphi}(y_1|x_1)-g^{(j)}_{\varphi}(y_2|x_2)|\leq C_x
(|y_1- y_2|^{\nu}+d(x_1, \ x_2)^{\beta}),\ j=0,2. $\\
\end{itemize}

\item [\bf{A4}] The kernel $H$ is  such that
\begin{enumerate}
        \item[(i)]$\displaystyle{\int_{\R}|t|^{\nu}H(t)dt<\infty}$ and
        $\displaystyle{\int_{\R}tH(t)dt=0}$,
        \item[(ii)] $\displaystyle{\forall\ (t_1,t_2)\! \in\!\R^2, |H(t_1\!-\!\varphi_1(z))\!-\!H(t_2\!-\!\varphi_2(z))|\!\leq \!C_H(|t_1\!-\!
t_2|\!+\!d_{\C}(\varphi_1,\varphi_2))}$, where $C_H$ is a positive
constant.
\end{enumerate}

 \item [\bf{A5}] For $j=0,1,2$ and any $\varphi\in {\cal C}$,
    $$\E\left\lbrack
H^{(j)}\left(\frac{y-\varphi(Z_i)}{h}\right)\mid\G_{i-1}\right\rbrack=\E\left\lbrack
H^{(j)}\left(\frac{y-\varphi(Z_i)}{h}\right)\mid\
X_i\right\rbrack.$$
\end{enumerate}

\noindent{\bf Comments on the hypotheses.}  As to discuss the conditions A1, it is worth noticing that the fundamental hypothesis {\bf A1}(ii) involves
the functional nature of the data together with their dependency. As usually in such a framework, small balls techniques are used to handle
the probabilities on infinite dimension spaces where the Lebesgue measure does not exist. Several examples of processes fulfilling this
condition are given in La\"ib \& Louani (2011). Note that the hypothesis {\bf A1}(i) stands as a particular case of {\bf A1}(ii) while conditioning
by the trivial $\sigma$-field. A number of processes satisfying this condition are given through out the literature, see, for instance, Ferraty \& Vieu (2006).
Conditions {\bf A1}(iii) and {\bf A1}(v) are set basically to meet the ergodic Theorem which may be expressed as the classical law of large numbers.
Conditions {\bf A2} and {\bf A4} impose some regularity
conditions upon the kernels used in our estimates.
When indexing by a class of functions $\C$, it is natural to consider regularity conditions as the continuity of the mode with respect to the index
function $\varphi$ assumed in {\bf A3}(i). Defined as an argmax and, furthermore, indexed by the class $\C$, the conditional mode is sensitive to
fluctuations. The diffrentiability of the conditional
density $g_\varphi$ with some kind of smoothness of its derivatives is needed to reach the rates of the convergence obtained in our results.
All these conditions are summarised in the assumption {\bf A3}.
Hypothesis {\bf A5} is of Markov's nature.

\vskip 5mm

Before  establishing the uniform convergence with rate, with
respect to the class of functions ${\cal C}$,  of the conditional
mode estimator, we introduce the following notation. For any
$\epsilon>0$, set
\begin{eqnarray*}
{\cal N}(\epsilon,{\cal C},d)&=&\min\{n:\ \mbox{there exist}\ c_1\cdots,c_n \ \mbox{in}\ {\cal C}\ \mbox{such that}\ \forall \ \varphi\in {\cal C}  \\
&& \hspace{1.3cm}\ \mbox{ there exists}\ 1\leq k\leq n \
\mbox{such that}\ d(\varphi,c_k)<\epsilon\}.
\end{eqnarray*}
This number measures how full is the class ${\cal C}$. Obviously,
conditions upon the number ${\cal N}(\epsilon,{\cal C},d)$ have to
be set to state uniform over the class ${\cal C}$ results.

\vskip 2mm
The following proposition establishes the uniform asymptotic behavior (with rate) of
the conditional density estimator $g(y | x)$ with respect to y and the function $\varphi \in {\cal C}$. This
proposition which is of interest by itself may be used, as an intermediate result, to prove
our main result given in Theorem 1 below.

\begin{proposition}\label{lemmasup}
Under conditions {\bf{A1}}, {\bf {A2}},  {\bf {A3}}, {\bf {A4}}
and {\bf {A5}} together with hypotheses
(\ref{eqnfirst})-(\ref{vc}), we have
\begin{eqnarray*}
\sup_{\varphi\in \C}\sup_{y\in
S_\varphi}|g_{\varphi,n}(y|x)-g_{\varphi}(y|x)|&=&O_{a.s}(h^\beta+h^\nu)+O_{a.s.}\left(\eta
h^{-2}\right)+O_{a.s.}\left(\lambda_n\right)\\
&&\ \ +O_{a.s}\left(\left(\frac{\log
n}{n\phi(h)}\right)^{1/2}\right).
\end{eqnarray*}
\end{proposition}


\vskip 2mm

Our principal result  considers the pointwise in $x\in{\cal E}$
and uniform over the class $\C$ convergence of  the kernel
estimate $\hat{\Theta}_{\varphi,n}(x)$ of the conditional mode
$\Theta_\varphi(x)$.

\begin{theorem}\label{th1}
Assume that the conditions {\bf {A1}}-{\bf {A5}} hold true and
that
\begin{eqnarray}\label{eqnfirst}
\lim_{n\rightarrow\infty} \frac{\log(n)}{n\phi(h)}=0.
\end{eqnarray}
Furthermore, for a sequence of positive real numbers $\lambda_n$
tending to zero, as $n\rightarrow\infty$, and
$\eta=\eta_n=O(h^{2+\beta})$, with $\beta>0$, suppose that
\begin{equation}\label{vc}
(i) \
 \ \lim_{n\rightarrow\infty}\frac{\log{\cal N}(\eta,{\cal
C},d)}{\eta\lambda_n^2nh\phi(h)}=0 \ \ \mbox{and}\ \ \ (ii) \ \
\sum_{n\geq 1}\exp\{-\lambda_n^2O(nh\phi(h))\}<\infty.
\end{equation}
Then, as $n\rightarrow\infty$, we have
$$\sup_{\varphi\in\C}|\hat{\Theta}_{\varphi,n}(x)-\Theta_\varphi(x)|=
O_{a.s}(h^{\beta/\!2}+h^{\nu/\!2})+O_{a.s}\left(\lambda_n^\frac{1}{2}\right)+O_{a.s}\left(\left(\frac{\log
n}{n\phi(h)}\right)^{1/4}\right).$$
\end{theorem}
\vspace{2mm}

\begin{rem}\
Replacing the condition (\ref{vc})(i) by
$\displaystyle{\lim_{n\rightarrow\infty}\frac{\log{\cal
N}(\eta,{\cal C},d)}{\eta\lambda_n^2nh\phi(h)}=\delta}$, for some
$\delta>0$, with $\displaystyle{\lambda_n=O\left(\left(\frac{\log
n}{nh\phi(h)}\right)^{\frac{1}{2}}\right)}$, the condition
(\ref{vc})(ii) is clearly satisfied inducing the uniform
consistency, with respect to $\varphi\in\C$, of
$\hat{\Theta}_{\varphi,n}(x)$ with the rate
$$\displaystyle{O_{\mbox{a.s.}}(h^{\beta/\!2}+h^{\nu/\!2})+O_{\mbox{a.s.}}\left(\left(\frac{\log
n}{n\phi(h)}\right)^{\frac{1}{4}}\right)}$$.
\end{rem}

\begin{rem}\  Note, whenever $\eta=O(h^{2+\beta})$ and
$\lambda_n^{-1}=O((nh^{5+2\beta}\phi(h))^{\frac{1}{2}})$, that the
condition (\ref{vc})(i) takes the form
\begin{equation}\label{vc1}
\lim_{\eta\rightarrow 0}\eta\log{\cal N}(\eta,{\cal C},d)=0.
\end{equation}
The condition (\ref{vc1}) is very usual in defining
Vapnik-Chervonenkis classes. Examples of classes fulfilling the
condition (\ref{vc1}) are given throughout the literature, see,
for instance, La\"ib {\rm \&} Louani (2011) and van der Vart {\rm
\&} Wellner (1996).
\end{rem}

\vspace{2mm}

\subsection{Application to time series forecasting}

The main application of Theorem 1 is devoted to prediction of time
series when considering the conditional mode estimates.

For $n\in \mathbb{N}^\star$, let $Z_i(t)$ and $X_i(t)$, $i=1, \dots, n$, be two functional random variables with $t\in [0,T).$
For each curve $X_i(t)$ (the covariate), we have a real response $Y_{\varphi,i} = \varphi(Z_i(t))$, a transformation of some
functional variable $Z_i(t)$. Given a new curve $X_{n+1}=x_{\mbox{new}}$, our purpose is to predict
the corresponding response $y_{\varphi,\mbox{new}} := \Theta_\varphi(x_{\mbox{new}})$ using as predictor the conditional mode, say
$\widehat{y}_{\varphi,\mbox{new}} := \widehat{\Theta}_{\varphi, n}(x_{\mbox{new}})$. The following Corollary
based on Theorem 1 gives the asymptotic behavior
with rate of the empirical error prediction.

\begin{corollary}\label{cor}
Assume that conditions of Theorem 1 hold. Then we have
$$
\Big| \widehat{y}_{\varphi,\mbox{new}} - y_{\varphi,\mbox{new}}\Big| = O_{a.s}(h^{\beta/\!2}+h^{\nu/\!2})+O_{a.s}\left(\lambda_n^\frac{1}{2}\right)+O_{a.s}\left(\left(\frac{\log
n}{n\phi(h)}\right)^{1/4}\right).
$$
\end{corollary}
\noindent {\it Proof.} The proof of Corollary \ref{cor} is a direct consequence of Theorem 1.



\vskip 5mm

\vspace{5mm}

\setcounter{section}{3}
\centerline{\large 3. APPLICATIONS TO REAL DATA}
\vspace{2mm}

\begin{figure}[ht!]
\begin{center}
\includegraphics[height=8cm,width=17cm]{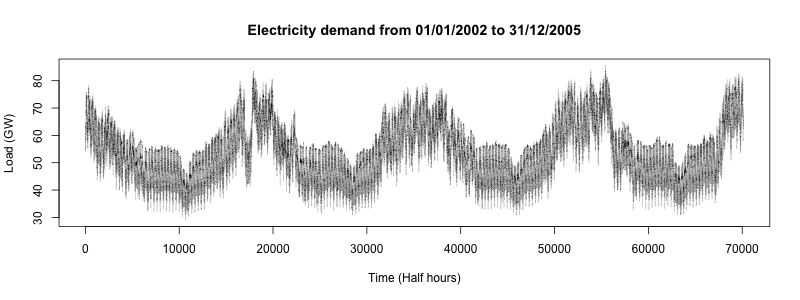}
\end{center}
\caption{Half hour electricity consumption in France from 07/01/2002 to 31/12/2005 generated according to the process $\xi(t)$. }
\label{load}
\end{figure}

\begin{figure}[ht!]
\begin{center}
\includegraphics[height=10cm,width=17cm]{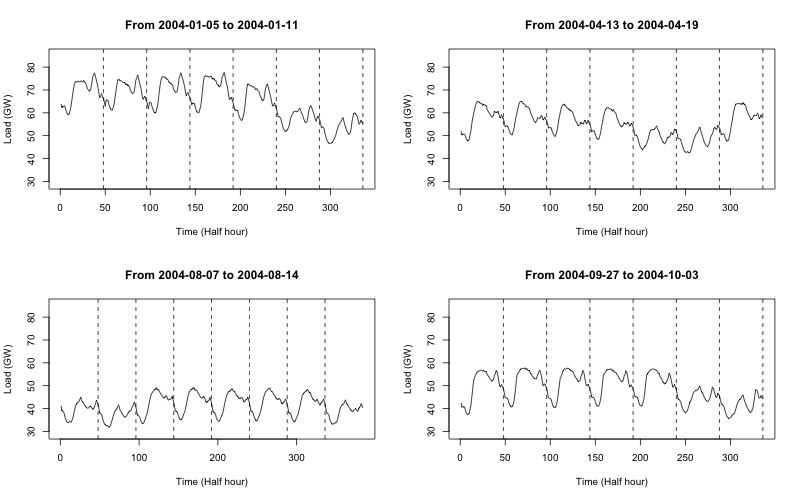}
\end{center}
\caption{Half hour electricity demand in four selected weeks (the panel contains one week data from January, April, August and October 2004). }
\label{load_saison}
\end{figure}

\begin{figure}[ht!]
\begin{center}
\includegraphics[height=7cm,width=10cm]{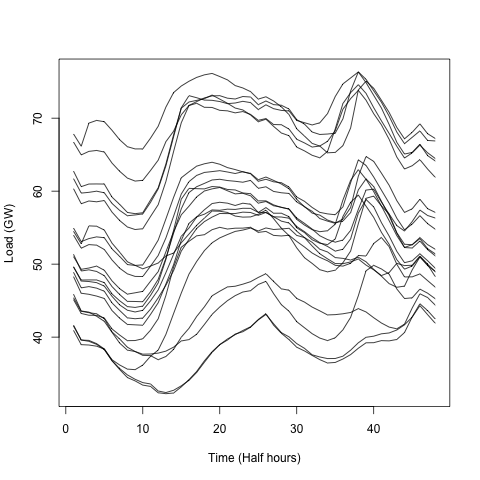}
\end{center}
\caption{A sample of 20 daily load curves randomly chosen generated according to the process $Z(t)$.}
\label{sample}
\end{figure}

The data-set analyzed in this paper contains half hourly observations of a stochastic process $\xi(t)$, $t\in \mathbb{R}^+$.
Here $\xi(t)$ represents the electricity demand at time $t$ in France. This process has been observed at each half hour
from 01 January 2002 to 31 December 2005 (which corresponds to a total of 1461 days). Figure \ref{load} shows the evolution of the process $\xi(t)$ over time.
One can easily see a high seasonality since the variation of the electricity consumption is due to
the climatic conditions in France. In fact, the winter and autumn are rather cold, whereas the climate in summer
and spring is relatively warm.
This remark is confirmed by Figure \ref{load_saison} which displays the half hourly electricity consumption
in France in four selected weeks. We can clearly distinguish the intra-daily periodical pattern, and we can also note the difference
in terms of level of consumption from one season to another. The repetitiveness of the daily shape is due to
a certain inertia in the demand that reflects the aggregated behavior of consumers.
Theoretically the evolution of the energy data was observed according to the process
$\xi(t)$. Now in order to construct our functional data $Z(t)$ and to get its transformation
$\varphi(Z(t))$ we proceed by slicing the original process $\xi(t)$ into segments of similar length. Since our target is a day-ahead short term forecasting we divide the observed original time series ($\xi(t)$) of half hourly electricity consumption into $n=1461$ segments (Z(t)) of length $48$ which correspond to the {\it functional observations}.
Each segment coincident with a specific daily load curve. Formally, let $[0,T]$ the time interval on which the process $\xi(t)$ is observed.
We divide this interval into subintervals of length 48, say $[\ell\times 48, (\ell+1)\times 48]$, $\ell = 0,1, \dots, n-1$, with $n=T/48 = 1461$.
Denoting by $Z_i(t)$ the functional-valued discrete-time stochastic process defined by
\begin{eqnarray}
Z_i(t) = \xi(t+(i-1)\times 48); \quad \quad i=1,\dots, n, \quad \forall t\in [0,48).
\end{eqnarray}
Figure \ref{sample} shows a, randomly chosen, sample of 20 realizations of the functional random variable $(Z(t))_{t\in [0,48)}$
which corresponds to a daily load curves.

Once we have transformed our original time series into functional data type, one can start, as explained in the introduction, dealing with the short term forecasting
of the daily peak demand and the electrical energy consumed over an interval using the conditional mode as predictor.

\subsection{Short-term daily peak load forecasting}
Let us now consider the observed daily peak of the electricity consumption
defined, for any day $i=1, \dots, n$, as
\begin{eqnarray*}\label{defsup}
\mathcal{P}_i = \sup_{t\in [0,48)} Z_i(t).
\end{eqnarray*}

The goal of this subsection is to forecast the peak $\mathcal{P}_i$ on the basis of the load curve of the previous day, $Z_{i-1}(t)$.
Forecasting peak load demand is one of the most relevant issues in electricity companies.
In fact, the electricity market is more and more open to competition
and companies take care on the quality of their services in order to increase the number of their customers.
On the other hand, because of the electrification of appliances (e.g. electric heating, air conditioning, \dots ) and mobility applications (e.g. electric vehicle, \dots), the peak demand is increasing which can leads to a serious issues in the electric network. It is important, therefore, to produce very accurate short-term peak demand forecasts for the day-to-day operation, scheduling and load-shedding plans of power utilities. Forecasting peak load toke a lot of interest in the statistical literature. For instance Goia {\it et al.} (2010) used a functional linear regression model and Sigauke \& Chikobvu (2010) a multivariate adaptive regression splines model. These methods are based on the regression function as a predictor, here we suggest the mode regression
as an alternative.

In this paper, we compare two predictors based on different choices of the covariable $X(t)$. Since the peak electricity demand is
highly correlated to the electricity consumption of the previous day and also to the temperature measures, we have then two possibilities to chose the covariable $X(t)$:
%
 \begin{figure}[ht!]
\begin{center}
\includegraphics[height=8cm,width=16cm]{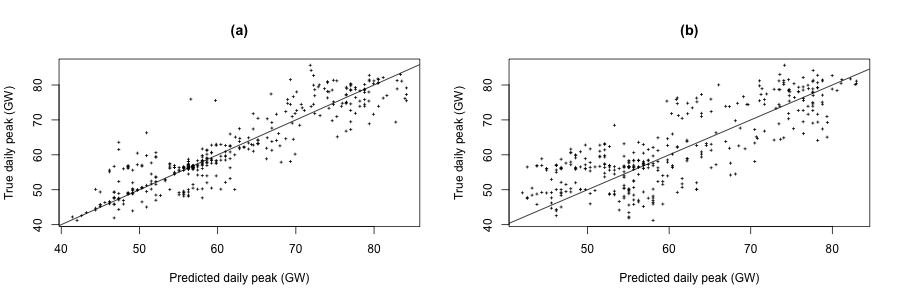}
\end{center}
\caption{Results for one year day-ahead forecasting of the daily peak electricity demand using as predictor the mode regression and as functional covariate: (a) the curve of electricity consumption of the previous day (\texttt{Prev.Day}), (b) the predicted half hourly temperature curve of the target day (\texttt{Pred.Temp.}).}
\label{result_peak}
\end{figure}
\begin{figure}[h!]
\begin{center}
\begin{tabular}{ll}
\includegraphics[height=8cm,width=8cm]{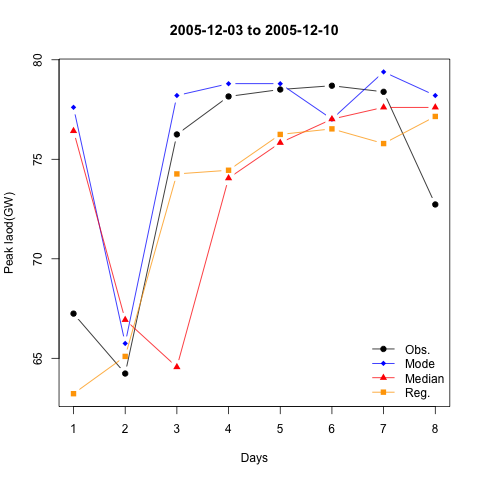} & \includegraphics[height=8cm,width=8cm]{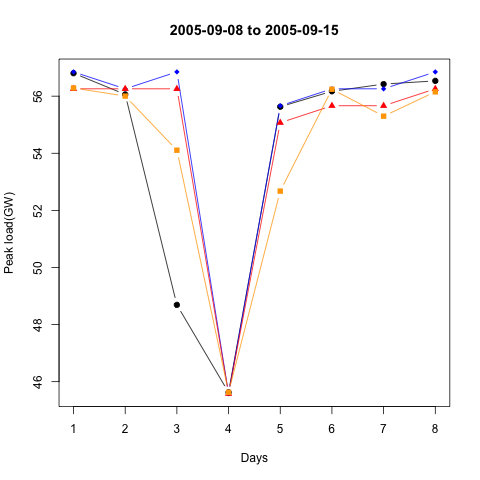}\\
\hspace{4cm}{\footnotesize{(a)}} & \hspace{4cm}{\footnotesize{(b)}}\\
\end{tabular}
\end{center}
\caption{Examples of daily peak load forecast using conditional mode, conditional median and regression function for eight consecutive days} 
\label{forecastpeak}
\end{figure}
%

To evaluate the proposed approach, we split the sample of $n=1461$ days into:

\begin{rem}
When the functional covariate is fixed to be the predicted temperature curve, say $\widehat{T}(t)$,
the notations for the learning and test sample
can be changed as follow:  $\mathcal{L}=\{(\widehat{T}_{i}(t), \mathcal{P}_{i}) \}_{i=1, \dots, 1096}$ and $\mathcal{T}= \{(\widehat{T}_{i}(t), \mathcal{P}_{i}) \}_{1, \dots, 365}$.
\end{rem}

\begin{table}[!t]
\renewcommand{\arraystretch}{1.3}
\caption{Distribution (by month) of the \texttt{RAE} of the peak load obtained by using as predictor the conditional mode and as covariate the previous day
\texttt{Prev.Day} and predicted temperature \texttt{Pred.Temp}.}
\label{table1}
\centering
\begin{tabular}{lcccc||cccc}
\hline
& &  \texttt{Prev.Day}(\%) & & & &   \texttt{Pred.Temp.}(\%) & &\\
   \hline
   \hline
 & $\texttt{MAPE}_m$ & $Q_{0.25}$ & $Q_{0.5}$ & $Q_{0.75}$ & $\texttt{MAPE}_m$ & $Q_{0.25}$ & $Q_{0.5}$ & $Q_{0.75}$   \\
\hline
\bfseries Jan.& 4.2 & 1.4 & 3.1 & 5.2 & 9.5 & 5.4 & 9.4 & 14.2\\
\bfseries Feb. & 5.4 & 1.9 & 4.0 & 8.3 & 6.9 & 1.3 & 6.3 & 10.2 \\
\bfseries Mar. & 6.1 & 2.0 & 3.7 & 8.0 & 9.0 & 3.5 & 6.5 & 13.3  \\
\bfseries Apr. & 6.7 & 2.6 & 4.9 & 9.4 & 11.9 & 6.0 & 11.4 & 18.2 \\
\bfseries May & 4.7 & 0.5 & 1.5 & 9.5 & 9.9 & 3.2 & 8.0 & 14.1 \\
\bfseries Jun. & 2.4 & 0.7 & 1.2 & 2.2 & 9.3 & 3.5 & 7.7 & 15.5\\
\bfseries Jul. & 3.1 & 0.5 & 1.5 & 3.2 & 8.3 &2.2 & 7.4& 11.2\\
\bfseries Aug. & 3.0 & 0.6 & 1.3 & 3.6 & 11.1 & 4.6 & 8.0 & 14.2\\
\bfseries Sep. & 2.5 & 0.3 & 0.8 & 1.7 & 10.9 & 4.0 & 9.5 & 18.7\\
\bfseries Oct. & 4.0 & 0.9 & 2.3 & 4.4 & 11.0 & 4.5 & 9.9 & 17.5\\
\bfseries Nov. & 5.4 & 2.0 & 3.9 & 7.1 & 8.9 & 3.7 & 6.8 & 14.1\\
\bfseries Dec. & 5.9 & 2.2 & 5.5 & 8.2 & 7.1& 3.2 & 5.8 & 9.3 \\
    \hline
\end{tabular}
\end{table}

The learning sample is used to build the proposed estimator given by (\ref{def_est})
and to find the ``optimal" smoothing parameter. To estimate the conditional mode some
tuning parameters should be fixed. We suppose here that the kernel (resp. the smoothing parameter) for both
the covariate and the response variable is the same and considered to be the quadratic kernel defined as
$K(u) = H(u)= 1.5(1-u^2)\1_{[0,1]}$ (resp. $h:= h_K=h_H$). The optimal
bandwidth $h$ is obtained by the cross-validation method on the $k$-nearest
 neighbors (see Ferraty and Vieu (2006), p. 102 for more details).
Finally, the semi-metric $d(\cdot, \cdot)$ is fixed to be the $L_2$ distance
between the second derivative of the curves.

 \vspace{3mm}

The test sample will be used to compare our forecasts to the observed daily peak electricity demand for
the year 2005. Figure \ref{result_peak}(a) (resp. (b)) displays the observed and the predicted values of
the daily peak electricity demand using as covariable the load curve of the previous day
(reps. the predicted temperature curve). Since cross-points, $(\widehat{\mathcal{P}}_i,\mathcal{P}_i)_{i=1, \dots, 365}$, represented in Figure \ref{result_peak}(a) are more concentrated on the diagonal line
than those in Figure \ref{result_peak}(b),
one can deduce that the first approach provides better results
than the second one. Moreover, Table \ref{table1} provides a numerical summary of the \texttt{RAE} obtained by using as covarite the predicted temperature curve or the last observed daily load curve. One can observe that monthly errors obtained by the second approach are usually less than those given by the first one. Therefore, one can conclude that the peak electricity demand might be better modelized by the
previous daily load curve rather than by the predicted temperature curve.
 \vspace{3mm}

 Following the previous analysis, the last observed daily load curve will be be considered, in the rest of this section, as the suitable covariate to forecast the peak demand. Our goal, now, consists in comparing the conditional mode predictor to the conditional median and the regression function (conditional mean) (see Ferraty and Vieu (2006) for more details about the properties of those last two predictors).

For a deeper analysis and evaluation of the accuracy of the proposed approach we use as validation criterias: the Relative Absolute
Errors (\texttt{RAE}) and the {\it monthly} Mean Absolute Prediction Error ($\texttt{MAPE}_m$), defined respectively, for any day $i = 1, \dots, 365$, in the test sample, as
$$
\texttt{RAE}_i = \frac{|\mathcal{P}_i - \widehat{\mathcal{P}}_i|}{\mathcal{P}_i}\quad\quad \mbox{and} \quad\quad \texttt{MAPE}_m = \frac{1}{N_m}\sum_{i=1}^{N_m} \frac{|\mathcal{P}_i - \widehat{\mathcal{P}}_i|}{\mathcal{P}_i}
$$

where $N_m$ is a number of days for a given month $m\in \{1, \dots, 12\}$ and $\widehat{\mathcal{P}}_i$ is the predicted value of the daily peak obtained by the conditional mode, conditional median or regression function.

\begin{figure}[ht!]
\begin{center}
\includegraphics[height=8cm,width=16cm]{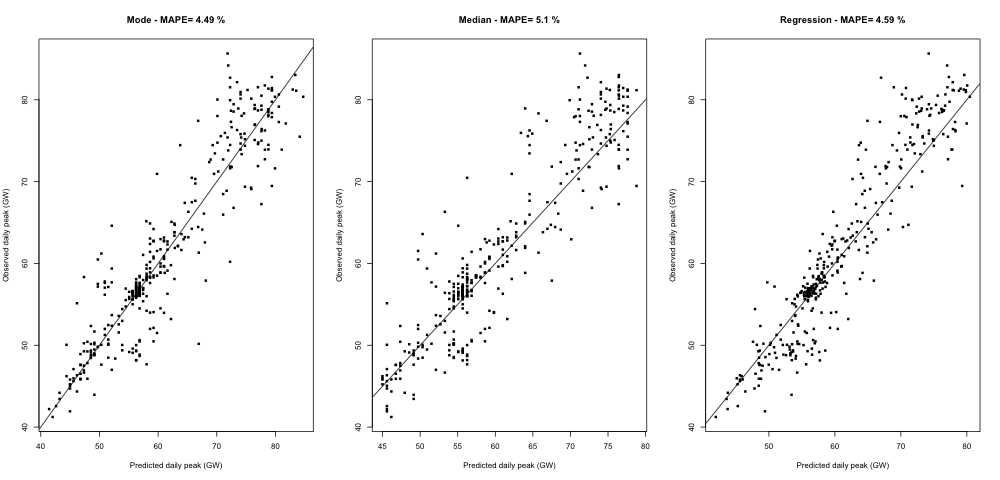}
\end{center}
\caption{Observed daily peak load versus the predicted one obtained by the three forecast methods.}
\label{for_three}
\end{figure}

Figure \ref{for_three} shows examples of peak load forecasts for eight consecutive days. One can see that conditional mode provides more accurate predictions than the two other methods. In Figure \ref{forecastpeak} the 365 forecasted daily peak load are plotted against the observed ones. Clearly, one can observe that conditional mode performs well the forecasts while conditional median and regression function under-predict peak in the cold season and over-predict it in hot season.

\begin{figure}[ht!]
\begin{center}
\includegraphics[height=8cm,width=16cm]{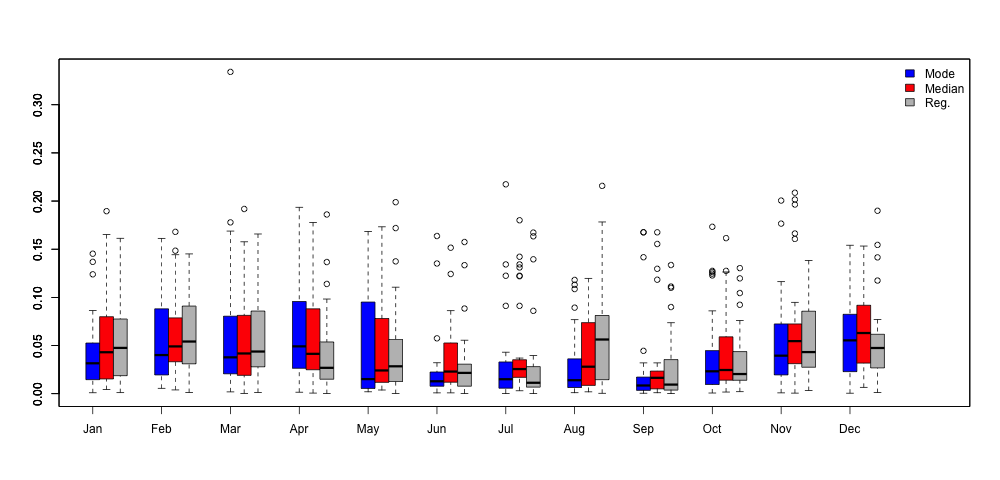}
\end{center}
\caption{Distribution (by month) of the daily \texttt{RAE} of the peak load.}
\label{boxplot_peak}
\end{figure}

Figure \ref{boxplot_peak} provides the distribution of the daily \texttt{RAE} for each month in 2005, obtained by using
the three prediction methods. One can observe that the conditional mode-based approach
is much more efficient in winter, as well as, in summer, than the other methods. Accurate forecasts in winter are of particular interest since the last is the period of the year whenever electricity demand might exceed the supply capacity and therefore efficient energy management in the electrical grid is highly required.

A numerical summary of Figure \ref{boxplot_peak} is detailed in Table \ref{table2} where the monthly $\texttt{MAPE}_m$, the first quartile $Q_{0.25}$, the median $Q_{0.5}$ and the third quartile $Q_{0.75}$, of the \texttt{RAE}, are provided for the three used methods. One can see that the conditional mode approach performs better the forecasts almost over all the year.


\begin{landscape}
\begin{table}[!t]
\renewcommand{\arraystretch}{1.3}
\caption{Distribution (by month) of the \texttt{RAE} of the peak load obtained by  using as covariate the previous day
\texttt{Prev.Day} and as prediction methods the conditional mode, conditional median and regression function respectively.}
\label{table2}
\centering
\begin{tabular}{lcccc||cccc||cccc}
\hline
& & \texttt{Mode}(\%)&    & & &  \texttt{Median}(\%)&   & & &  \texttt{Reg.}(\%) & & \\
   \hline
   \hline
 & $\texttt{MAPE}_m$& $Q_{0.25}$ & $Q_{0.5}$ & $Q_{0.75}$ & $\texttt{MAPE}_m$& $Q_{0.25}$ & $Q_{0.5}$ & $Q_{0.75}$ &$\texttt{MAPE}_m$& $Q_{0.25}$ & $Q_{0.5}$ & $Q_{0.75}$   \\
\hline
\bfseries Jan.& 4.2& 1.4 & 3.1 & 5.2 & 5.8 & 1.5 & 4.3 & 7.9 & 5.1& 1.8 & 4.7 & 7.7\\
\bfseries Feb. & 5.4& 1.9 & 4.0 & 8.3 & 5.9& 3.3 & 4.9 & 7.8 &6.4&  3.1 & 5.4 & 9.0 \\
\bfseries Mar. & 6.1& 2.0 & 3.7 & 8.0 & 5.6&  1.9 & 4.1 & 8.1 & 6.0 & 2.7 & 4.3 & 8.5  \\
\bfseries Apr. &6.7&  2.6 & 4.9 & 9.4 & 5.9& 2.5 & 4.1 & 8.4 & 4.1 & 1.5 & 2.6 & 5.2 \\
\bfseries May & 4.7& 0.5 & 1.5 & 9.5 & 4.6& 1.1 & 2.4 & 7.8 &4.5 &  1.2 & 2.8 & 5.6  \\
\bfseries Jun. & 2.4&  0.7 & 1.2 & 2.2 & 3.7& 1.2 & 2.2 & 5.2 &2.9 & 0.8 & 2.1 & 3.0 \\
\bfseries Jul. &3.1& 0.5 & 1.5 & 3.2 & 4.5&  1.6 & 2.5 & 3.5 &2.9 & 0.6 & 1.1 & 2.8 \\
\bfseries Aug. & 3.0& 0.6 & 1.3 & 3.6 & 4.0& 0.8 & 2.8 & 7.3 & 5.8 &1.4 & 5.6 & 8.1 \\
\bfseries Sep. & 2.5& 0.3 & 0.8 & 1.7 & 3.0&  0.5 & 1.6 & 2.3 &2.7 & 0.4 & 0.9 & 3.1\\
\bfseries Oct. & 4.0& 0.9 & 2.3 & 4.4 & 4.2& 1.4 & 2.4 & 5.9 &3.4 & 1.3 & 2.0 & 4.3 \\
\bfseries Nov. & 5.4&  2.0 & 3.9 & 7.1 & 6.8 & 3.1 & 5.4 & 7.1 &5.3 & 2.8 & 4.3 & 8.3\\
\bfseries Dec. & 5.9& 2.2 & 5.5 & 8.2 & 6.6 & 3.1 & 6.2& 9.1 &5.4 & 2.6 & 4.7 & 6.1 \\
    \hline
\end{tabular}
\end{table}
\end{landscape}

\subsection{Electrical energy consumption forecasting for battery storage management}
The electrical grid in the majority of the developed countries is expected to be put under a large
amount of strain in the future due to changes in demand behavior, the electrification of transport,
heating and an increased penetration of distributed generation. The current electrical grids
 infrastructure may not be able to endure these changes and storage of energy, produced
 by solar and wind power generations, hopes to negate or postpone the need for expensive
 conventional reinforcement that may be needed due to these changes in demand.
 One of the most used approaches to solve this technical issue consists in the storage, in batteries, of the energy
 coming from the traditional energy plants (e.g. nuclear, hydraulic, \dots ) and from renewable
 energy resources (e.g. solar and wind) during the day and then use it at the evening and especially
 over the three hours around the peak (around 7pm in winter and 2pm in summer). Therefore, an accurate
 forecast of the energy that will be consumed in the evening allows to optimize the capacity of the storage and
 consequently to increase the batteries' life.

  \vspace{3mm}

\begin{figure}[h!]
\begin{center}
\begin{tabular}{ll}
\includegraphics[height=8cm,width=8cm]{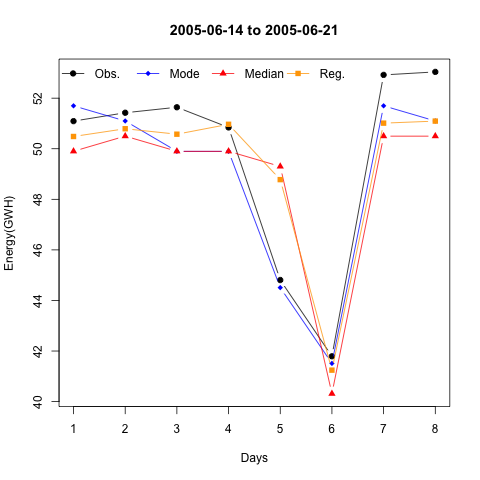} & \includegraphics[height=8cm,width=8cm]{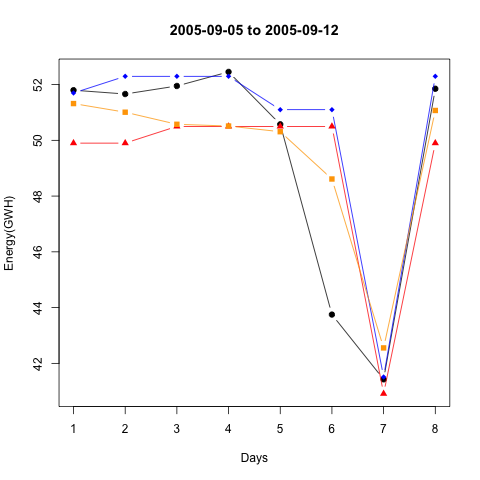}\\
\hspace{4cm}{\footnotesize{(a)}} & \hspace{4cm}{\footnotesize{(b)}}\\
\end{tabular}
\end{center}
\caption{Examples of energy demand forecast using conditional mode, conditional median and regression function for eight consecutive days} 
\label{forecastenergy}
\end{figure}

\begin{figure}[ht!]
\begin{center}
\includegraphics[height=8cm,width=17cm]{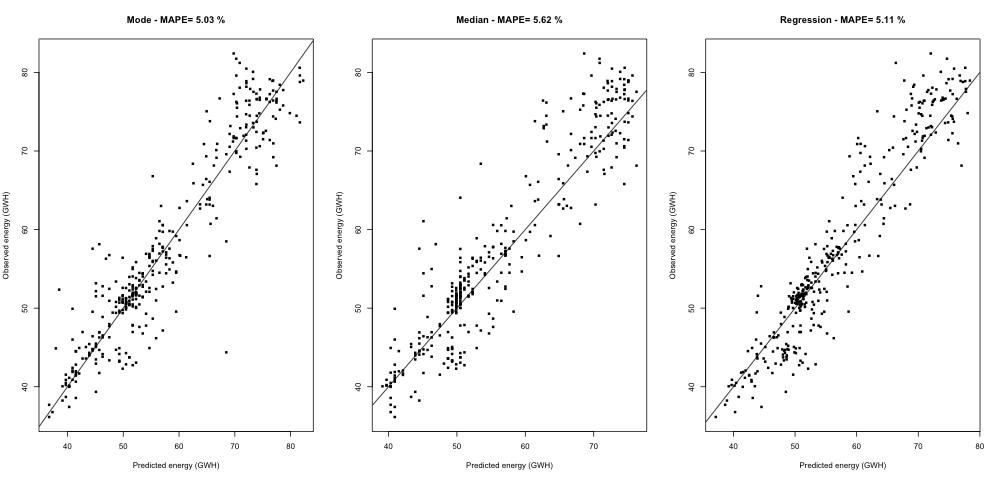}
\end{center}
\caption{Observed daily consumed energy in 2005 versus its predicted values.}
\label{boxplot_three_mean}
\end{figure}

\begin{figure}[ht!]
\begin{center}
\includegraphics[height=8cm,width=16cm]{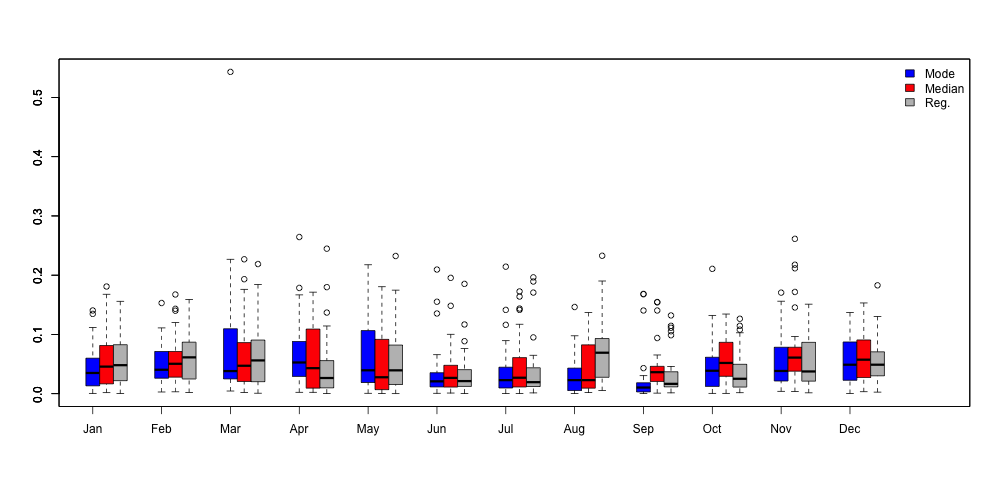}
\end{center}
\caption{Distribution (by month) of the daily \texttt{RAE} of the consumed energy.}
\label{boxplot_energy}
\end{figure}

 In this subsection, we suggest to solve this forecasting issue by using the mode regression.
 Regarding to the discussion made in the previous subsection, we use as covariate the load curve of
 the previous day. Formally, if we consider $Z_i(t)$ the load curve of some day $i$,
 then the electrical energy consumed between $t_1$ and $t_2$
is defined as $\mathcal{E}_i = \int_{t_1}^{t_2} Z_i(t) dt$ and measured in Giga-Watt per Hour (GWH).
Therefore, $\varphi(\cdot)$ in that case is the integral function. Here, we use the same data set and the same evaluation
procedure used in the previous subsection. We also keep the same choices for the tuning parameters ($K$, $H$, $d(\cdot, \cdot$) and $h$) in the model. As mentioned before, the functional covariable is supposed to be the last observed daily load curve. Figure \ref{forecastenergy} and \ref{boxplot_three_mean} show that conditional mode approach performs energy forecasts and that conditional median, as well as, regression function, under-predict energy for cold days and over-predict it in hot ones. Figure \ref{boxplot_energy} provides the distribution, by month, of the daily \texttt{RAE}
 and we can observe that accurate results are obtained with conditional mode predictor. Numerical details, namely monthly \texttt{MAPE}, first quartile $Q_{0.25}$, the median $Q_{0.5}$ and the third quartile $Q_{0.75}$, of the obtained errors are given in Table \ref{table3}. One can see again that conditional mode performs better the forecasts of the consumed energy

\begin{figure}[ht!]
\begin{center}
\includegraphics[height=6cm,width=11cm]{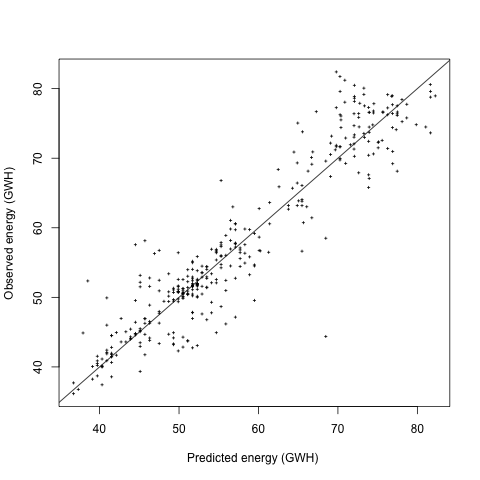}
\end{center}
\caption{Observed daily consumed energy in 2005 versus its predicted values.}
\label{scatterplot}
\end{figure}
%


\begin{landscape}
\begin{table}[!t]
\renewcommand{\arraystretch}{1.3}
\caption{Distribution (by month) of the relative absolute errors of the energy consumed obtained by  using as covariate the previous day
\texttt{Prev.Day} and as prediction methods the conditional mode, conditional median and regression function respectively.}
\label{table3}
\centering
\begin{tabular}{lcccc||cccc||cccc}
\hline
& & \texttt{Mode}(\%)&    & & &  \texttt{Median}(\%) &   & &&  \texttt{Reg.}(\%)&   & \\
   \hline
   \hline
 &   $\texttt{MAPE}_m$ & $Q_{0.25}$ & $Q_{0.5}$ & $Q_{0.75}$ &  $\texttt{MAPE}_m$ & $Q_{0.25}$ & $Q_{0.5}$ & $Q_{0.75}$ & $\texttt{MAPE}_m$ & $Q_{0.25}$ & $Q_{0.5}$ & $Q_{0.75}$   \\
\hline
\bfseries Jan.& 4.3 & 1.3 & 3.4 & 5.9 & 5.7 & 1.6 & 4.5 & 8.1 & 5.4 & 2.2 & 4.8 & 8.2 \\
\bfseries Feb. & 5.0 & 2.6 & 4.0 & 5.0 &5.7 &  2.7 & 5.0 & 6.9 &6.2 & 2.6 & 6.1 & 8.2 \\
\bfseries Mar. & 8.6 &  2.4 & 3.8 & 10.9 & 6.3 & 2.0 & 4.7 & 8.6 & 6.6 & 2.0 & 5.6 & 9.0  \\
\bfseries Apr. & 6.8 & 3.0 & 5.2 & 8.8 & 6.4 & 1.3 & 4.3 & 10.3 & 4.7 & 1.1 & 2.6 & 5.5 \\
\bfseries May & 6.4 & 1.8 & 3.9 & 10.6 & 5.2 & 0.6 & 2.7 & 9.1 & 5.7 & 1.5 & 3.9 & 8.2  \\
\bfseries Jun. & 3.6 & 1.1 & 2.0 & 3.5 & 4.0 & 1.1 & 2.6 & 4.7 & 3.5 & 1.2 & 2.1 & 3.9 \\
\bfseries Jul. & 3.9&  0.9 & 2.2 & 4.4 & 4.8 & 1.1 & 2.6 & 6.0 & 4.0 & 1.2 & 1.9 & 4.3 \\
\bfseries Aug. & 3.2 & 0.5 & 2.2 & 4.3 & 4.6 & 0.9 & 2.2 & 8.2 & 6.9 & 2.7 & 6.9 & 9.3 \\
\bfseries Sep. & 2.5 & 0.3 & 1.0 & 1.7 & 4.4 & 2.1 & 3.6 & 4.5 & 3.3 & 1.1 & 1.6 & 3.5\\
\bfseries Oct. & 4.5 & 1.2 & 3.8 & 6.1 & 5.5 & 2.9 & 5.1 & 8.6 & 3.7 & 1.1 & 2.5 & 4.9 \\
\bfseries Nov. & 5.2 & 2.1 & 3.8 & 7.7 & 7.6 & 3.7 & 6.0 & 7.7 & 5.3 & 2.2 & 3.7 & 8.4\\
\bfseries Dec. & 5.8 & 2.2 & 4.8 & 8.7 & 6.5 & 2.7 & 5.7 & 9.0 & 5.4 & 3.0 & 4.8 & 7.0 \\
    \hline
\end{tabular}
\end{table}
\end{landscape}

\clearpage
\vspace{5mm}

\setcounter{section}{2}
\centerline{\large 4. PROOFS}
\vspace{2mm}

In order to proof our results, we introduce some further notation.
 Let
\begin{eqnarray*}
 \bar f_{\varphi,n}(x,\
y) &=& \frac{1}{nh\E\lbrack \Delta_1(x)\rbrack}\sum_{i=1}^{n}
\E\left\lbrack \Delta_i(x)H\left(\frac{y-\varphi(Z_i)}{h}\right)
\mid \F_{i-1}\right\rbrack.
\end{eqnarray*}
and
$$\bar{l}_{n}(x)=\frac{1}{n\mathbb{E}(\Delta_1(x))}\sum_{i=1}^n\mathbb{E}\left(\Delta_i(x) | {\cal F}_{i-1}\right).$$
Define the conditional bias of the conditional density estimate of
$\varphi(Z_i)$ given $X=x$ as
$$B_{\varphi,n}(x,y)=\frac{\bar{f}_{\varphi,n}(x,y)}{\bar{l}_{n}(x)}-g_{\varphi}(y|x).$$
Consider now the following quantities
$$R_{\varphi,n}(x,y)=-B_{\varphi,n}(x,y)({l}_{n}(x)-\bar{l}_{n}(x)),$$ and
$$Q_{\varphi,n}(x,y)=({f}_{\varphi,n}(x,y)-\bar{f}_{\varphi,n}(x,y))-g_{\varphi}(y|x)({l}_{n}(x)-\bar{l}_{n}(x)).$$
It is then clear that the following decomposition holds
\begin{eqnarray}\label{conditional.density.1}
g_{\varphi,n}(y|x)-g_{\varphi}(y|x)=B_{\varphi,n}(x,y)+\frac{R_{\varphi,n}(x,y)+Q_{\varphi,n}(x,y)}{{l}_{n}(x)}.
\end{eqnarray}

The proofs of our results need the following lemmas as tools for
which details of their proofs may be found in La\"ib and Louani
(2011).

\begin{lemma}\label{lemma0}
Let $(X_n)_{n\geq 1}$  be a sequence of martingale differences
with respect to the sequence of $\sigma$-fields $({\cal
F}_n=\sigma(X_1,\cdots,X_n))_{n\geq 1}$, where
$\sigma(X_1,\cdots,X_n)$ is the $\sigma$-field generated by the
random variables $X_1,\cdots,X_n$. Set $S_n=\sum_{i=1}^nX_i$.
Suppose that the random variables $(X_i)_{i\geq 1}$ are bounded by
a constant $M>0$, i.e., for any $i\geq 1,\ \  |X_i|\leq M$ almost
surely, and $\mathbb{E}(X_i^2|{\cal F}_{i-1})\leq d_i^2$ almost
surely. Then  we have, for any $\lambda>0$, that
$$\mathbb{P}(|S_n|>\lambda)\leq2\exp\left\{-\frac{\lambda^2}{4D_n+2M\lambda}\right\},$$
where $D_n=\sum_{i=1}^nd_i^2.$
\end{lemma}

\vskip 4mm
\begin{lemma}\label{lemme1}
Assume that conditions  {\bf {A1}} \textnormal{((i), (ii), (iv))}
and  {\bf {A2}} hold true. For $1\leq j\leq2+\delta$ for some
$\delta>0$, we have
\begin{eqnarray*}
    (i) &&\frac{1}{\phi(h)}\E\lbrack\Delta^{j}_{i}(x)|\F_{i-1}\rbrack=M_{j}f_{i,1}(x)+O_{a.s}\left(\frac{\psi_{i,x}(h)}{\phi(h)}\right),\\
    (ii) &&\frac{1}{\phi(h)}\E\lbrack\Delta^{j}_{1}(x)\rbrack=M_{j}f_{1}(x)+o(1),\\
\end{eqnarray*}
where
$\displaystyle{M_j=K^j(1)-\int_{0}^{1}(K^j)^{\prime}\tau_0(u)du}$.
\end{lemma}
\vspace{3mm}
%
%
%

\noindent{\bf {Proof of Proposition}} \ref{lemmasup}. 
Considering the decomposition (\ref{conditional.density.1}), the
proof follows from lemmas \ref{lemmeln}, \ref{lemme3},   \ref{lem3} and \ref{lemme4} given hereafter, establishing
respectively the convergence of $l_{n}(x)$ to $1$ together with
the rate convergence of $l_n(x)-\bar{l}_n(x)$ to zero and the
orders of terms $B_{\varphi,n}(x,y)$, $R_{\varphi,n}(x,y)$ and
$Q_{\varphi,n}(x,y)$. Note that, due to the condition
(\ref{eqnfirst}), the term $R_{\varphi,n}(x,y)$ is negligible as
compared to the term $B_{\varphi,n}(x,y)$.
\hfill$\Box$\vspace{5mm}

\begin{lemma}\label{lemmeln}
Under assumptions {\bf {A1}} and {\bf {A2}}, we have
\begin{eqnarray*}
(i) && l_n(x)-\bar{l}_n(x)=O_{a.s}\left(\sqrt{\frac{\log(n)}{n\phi(h)}}\right),\\
(ii)&&
\lim_{n\rightarrow\infty}l_n(x)=\lim_{n\rightarrow\infty}\bar{l}_n(x)=1,
\quad a.s.
\end{eqnarray*}
\end{lemma}
\noindent{\it Proof of Lemma \ref{lemmeln}.} The results follow by
making use of Lemma \ref{lemma0} and Lemma 2  in La¨\"ib \& Louani (2011). Details
of the proof may be found in La\"ib and Louani (2010).\hfill$\Box$
\vspace{3mm}

\begin{lemma}\label{lemme3}
Under assumptions {\bf {A1}}, {\bf {A2}}, {\bf
{A3}}\textnormal{(iv)}, {\bf {A4}}\textnormal{(i)} and {\bf {A5}},
we have,  as $ n\rightarrow\infty$,
\begin{eqnarray*}
\sup_{\varphi\in\C}\sup_{y\in S_{\varphi}}(\bar f_{\varphi,n}(x,\
y)-\bar{l}_n(x)g_{\varphi}(y|x))=O_{a.s.}(h^{\beta}+h^{\nu}).\end{eqnarray*}
\end{lemma}
\noindent{\it Proof of Lemma \ref{lemme3}.} \ \ By condition
{\bf{A5}} with $j=0$, we have
\begin{eqnarray*}
  \bar f_{\varphi,n}(x,\ y)
   &=& \frac{1}{nh\E\lbrack \Delta_1(x)\rbrack}\sum_{i=1}^{n} \E\left\lbrack
\Delta_i(x)\E\left(H\left(\frac{y-\varphi(Z_i)}{h}\right)\mid
X_i\right)|\F_{i-1}\right\rbrack.
\end{eqnarray*}
A change of variables and the fact that
$\displaystyle{\int_{\R}H(t)dt}=1$ allow us to write
\begin{eqnarray*}
  \E\left\lbrack H\left(\frac{y-\varphi(Z_i)}{h}\right)\mid\ X_i\right\rbrack
  &=&h\int_{\R}H(t)\lbrack
  g_{\varphi}(y-th|X_i)-g_{\varphi}(y|x)\rbrack dt+h
  g_{\varphi}(y|x) \\
  &=:& J_{i,\varphi}+hg_{\varphi}(y | x).
\end{eqnarray*}
Thus,
\begin{eqnarray*}
\bar f_{\varphi,n}(x,y)\!
   &\!=\!&\!\frac{g_{\varphi}(y|x)}{n\E\lbrack \Delta_1(x)\rbrack}\!\sum_{i=1}^{n}
\!\E\left\lbrack \Delta_i(x)|\F_{i-1}\right\rbrack
+\frac{1}{nh\E\lbrack \Delta_1(x)\rbrack}\!\sum_{i=1}^{n}
\!\E\left\lbrack
\Delta_i(x)J_{i,\varphi}|\F_{i-1}\right\rbrack\\
 &=:& \bar{l}_{n}(x)g_{\varphi}(y|x)+S_{2}.\end{eqnarray*}
Using  condition {\bf A3}(iv), one may write
 \begin{eqnarray*}
|S_{2}|&\leq&\frac{1}{nh\E\lbrack
\Delta_1(x)\rbrack}\sum_{i=1}^{n} \E\left\lbrack
\Delta_i(x)|J_{i,\varphi}||\F_{i-1}\right\rbrack\\
&\leq&\frac{1}{n\E\lbrack \Delta_1(x)\rbrack}\sum_{i=1}^{n}
\E\left\lbrack \Delta_i(x)\left\{ \int_{\R} H(t)
\left(|t|^{\nu}h^{\nu}+(d(x,\
X_i))^{\beta}\right)dt\right\}|\F_{i-1}\right\rbrack\\
 &\leq&C_x\left\{h^{\nu} \int_{\R}|t|^{\nu} H(t)dt+h^{\beta}\right\}\frac{1}{n\E\lbrack \Delta_1(x)\rbrack}\sum_{i=1}^{n} \E\left\lbrack
\Delta_i(x)|\F_{i-1}\right\rbrack.
\end{eqnarray*}
Moreover,  considering Lemma 2 in La\"ib \& Louani (2011) combined with  the condition {\bf
A4}(i) imply that
\begin{eqnarray*}
\bar f_{\varphi,n}(x,\
y)-\bar{l}_n(x)g_{\varphi}(y|x)=O_{a.s.}(h^{\beta}+h^{\nu}),\end{eqnarray*}
where $O_{a.s.}$ does not depend on $\varphi\in \C$. \hfill$\Box$
\vspace{3mm}


The following Lemma describes the asymptotic behavior of the
conditional bias term $B_{\varphi,n}(x,y)$ as well as that of
$R_{\varphi,n}(x,y)$ and $Q_{\varphi,n}(x,y)$.
\begin{lemma}\label{lem3}
Under conditions {\bf{A1}}, {\bf {A2}},  {\bf {A3}} (ii), {\bf
{A4}} (i) and {\bf {A5}}, we have
\begin{equation}\label{biais}
\sup_{\varphi\in\C}\sup_{y\in
S_\varphi}|B_{\varphi,n}(x,y)|=O_{\mbox{a.s.}}(h^\beta+h^\nu),
\end{equation}
\begin{equation}\label{reste}
\sup_{\varphi\in\C}\sup_{y\in
S_\varphi}|R_{\varphi,n}(x,y)|=O_{\mbox{a.s.}}\left((h^\beta+h^\nu)\left(\sqrt{\frac{\log(n)}{n\phi(h)}}\right)\right).
\end{equation}
Moreover, when hypotheses (\ref{eqnfirst})-(\ref{vc}) are
satisfied, we have
\begin{equation}\label{Q}
\sup_{\varphi\in\C}\sup_{y\in
S_\varphi}|Q_{\varphi,n}(x,y)|=O_{a.s.}\left(\eta
h^{-2}\right)+O_{a.s.}\left(\lambda_n\right)+O_{a.s}\left(\left(\frac{\log
n}{n\phi(h)}\right)^{1/2}\right).
\end{equation}
\end{lemma}

\noindent{\it Proof of Lemma \ref{lem3}}.\ \ Observe that
$$B_{\varphi,n}(x,y)=\frac{\overline{f}_{\varphi,n}(x,y)-g_{\varphi}(y|x)\overline{l}_{n}(x)}{\overline{l}_{n}(x)}:=\frac{\tilde{B}_{\varphi,n}(x,y)}{\overline{l}_{n}(x)}.$$
Making use of  Lemma \ref{lemme3}, we obtain
$\sup_{\varphi\in\C}\sup_{y\in
S_\varphi}|\tilde{B}_{\varphi,n}(x,y)|=O_{\mbox{a.s.}}(h^{\beta}+h^\nu).$
The statement (\ref{biais}) follows  then from  the second part of
Lemma \ref{lemmeln}.

To deal now with the quantity $R_{\varphi,n}(x,y)$, write it as
$$R_{\varphi,n}(x,y)=-\frac{\tilde{B}_{\varphi,n}(x,y)}{\overline{l}_{n}(x)}(l_n(x)-\bar{l}_n(x)).$$
Therefore, the statement (\ref{reste}) follows from the statement
(\ref{biais}) combined with  Lemma
 \ref{lemmeln} (i).

In order to check the result (\ref{Q}), recall that
$$Q_{\varphi,n}(x,y)=({f}_{\varphi,n}(x,y)-\bar{f}_{\varphi,n}(x,y))-g_{\varphi}(y|x)({l}_{n}(x)-\bar{l}_{n}(x)).$$
Therefore  the statement (\ref{Q}) results from  Lemma
\ref{lemmeln} and  the use of Lemma \ref{lemme4} established
hereafter. This completes the proof of Lemma \ref{lem3}.
\hfill$\Box$ \vspace{3mm}

The following Lemma is needed as a step in proving Theorem
\ref{th1}

\begin{lemma}\label{lemme4}
Under assumptions {\bf {A1}}, {\bf {A2}}, {\bf {A3}}, {\bf
{A4}}\textnormal{(ii)}, {\bf {A5}} together with hypotheses
(\ref{eqnfirst})-(\ref{vc}), for $ n$ large enough, we have
\begin{eqnarray*}
\sup_{\varphi\in\C}\sup_{y\in S_\varphi}|f_{\varphi,n}(x,\
y)-\bar{f}_{\varphi,n}(x,\ y)|=O_{a.s.}\left(\eta
h^{-2}\right)+O_{a.s.}\left(\lambda_n\right).
\end{eqnarray*}
\end{lemma}

\noindent{\it Proof of Lemma \ref{lemme4}}.  Recall that, for any
$\varphi \in{\cal C}$,
$$S_{\varphi}=[ \theta_{\varphi}(x)-\xi, \  \ \theta_{\varphi}(x)+\xi].$$
Let $\varphi_1, \varphi_2\in\C$ and, for any $\epsilon>0$, define
the set
$$S_{\varphi_2}^\epsilon=\{ y\in\R :
|y-\Theta_{\varphi_2}(x)|\leq\xi+\epsilon\}.$$ It is easily seen,
by  condition {\bf A3} (i),
 that, for any $\epsilon>0$, there exists $\eta>0$ for which the fact that  $\varphi_1\in B(\varphi_2, \eta)$ implies
$S_{\varphi_1}\subset S_{\varphi_2}^\epsilon$.  Therefore, we have
\begin{eqnarray}\label{decompo.2}
\lefteqn{\sup_{\varphi\in\C}\sup_{y\in
S_\varphi}|f_{\varphi,n}(x,\ y)-\bar{f}_{\varphi,n}(x,\
y)|}\nonumber\\
&\hspace*{1.5cm}\leq&\max_{1\leq j\leq
\N(\eta,\C,d_{\C})}\sup_{\varphi\in B(\varphi_j,\eta)}\sup_{y\in
S_{\varphi}}|f_{\varphi,n}(x,\
y)-\bar{f}_{\varphi,n}(x,\ y)| \nonumber\\
&\hspace*{1.5cm}\leq& \max_{1\leq
j\leq\N(\eta,\C,d_{\C})}\sup_{\varphi\in
B(\varphi_j,\eta)}\sup_{y\in
S^\epsilon_{\varphi_j}}|f_{\varphi,n}(x,\
y)-\bar{f}_{\varphi,n}(x,\ y)|.
\end{eqnarray}
Using now the compactness of $S^\epsilon_{\varphi_j}$ and the fact
that its length is $2(\xi+\epsilon)$ for any $\varphi_j$, we can
write
$S^\epsilon_{\varphi_j}\subset\cup_{k=1}^{d_{\epsilon,n}}S^\epsilon_{\varphi_j,k}$
where $S^\epsilon_{\varphi_j,k}=(t^\epsilon_{\varphi_j,k}-m_n;\
t^\epsilon_{\varphi_j,k}+m_n)$ and $m_n$ and $d_{\epsilon, n}$ are
such that $ d_{\epsilon,n}=C_{\epsilon}m_n^{-1}$ for some positive
constant $C_{\epsilon}$. Moreover, we have
\begin{eqnarray}\label{decompostion.2}
\sup_{y\in S^\epsilon_{\varphi_j}}|f_{\varphi,n}(x,\
y)-\bar{f}_{\varphi,n}(x,\ y)| &\leq& \max_{1\leq k\leq
d_{\epsilon, n}}\sup_{y\in S^\epsilon_{\varphi_j,k}}|
f_{\varphi,n}(x,y)-f_{\varphi,n}(x,t^\epsilon_{\varphi_j,k}
)|\nonumber\\
&&+\max_{1\leq k\leq d_{\epsilon, n}}|
f_{\varphi,n}(x,t^\epsilon_{\varphi_j,k})-\bar{f}_{\varphi,n}(x,t^\epsilon_{\varphi_j,k})|
  \nonumber \\
&&+ \max_{1\leq k\leq d_{\epsilon, n}}\sup_{y\in
S^\epsilon_{\varphi_j,k}}|
\bar{f}_{\varphi,n}(x,t^\epsilon_{\varphi_j,k})-\bar{f}_{\varphi,n}(x,y)|\nonumber\\&=:&
J_{\varphi,n,1}+J_{\varphi,n,2}+J_{\varphi,n,3}.
\end{eqnarray}

Making use of {\bf A4} (ii), we obtain
\begin{eqnarray}\label{Jn1}
J_{\varphi,n,1}\!&\!\leq\!&\!\frac{1}{nh\E\lbrack
\Delta_1(x)\rbrack}\!\max_{1\leq k\leq d_{\epsilon,
n}}\!\sup_{y\in S^\epsilon_{\varphi_j,k}}\!\sum_{i=1}^{n}\!
\Delta_i(x)\!\left|H\!\!\left(\!\frac{y\!-\!\varphi(Z_i)}{h}\!\right)\!-\!H\!\!\left(\!\frac{t^\epsilon_{\varphi_j,k}\!-\!\varphi(Z_i)}{h}\!\right)
\!\right| \nonumber\\ &\leq &
C_Hm_nh^{-2}l_n(x).
\end{eqnarray}
Similarly, we  have also
\begin{eqnarray}\label{Jn3}
J_{\varphi,n,3}&\leq& C_Hm_nh^{-2}\bar{l}_n(x).
\end{eqnarray}
Therefore,
\begin{eqnarray}\label{J23}\max_{1\leq j\leq
\N(\eta,\C,d_{\C})}\sup_{\varphi\in
B_{(\varphi_j,\eta)}}\left(J_{\varphi,n,1}+J_{\varphi,n,3}\right)
\leq C_Hm_nh^{-2}\left(l_n(x)+\bar{l}_n(x)\right).\end{eqnarray}
Using Lemma \ref{lemmeln} (ii), it follows that
\begin{eqnarray}\max_{1\leq j\leq
\N(\eta,\C,d_{\C})}\sup_{\varphi\in
B(\varphi_j,\eta)}\left(J_{\varphi,n,1}+J_{\varphi,n,3}\right)
&=&O_{a.s.}\left(m_nh^{-2}\right).
\end{eqnarray}
To identify the convergence rate to zero of the term
$\displaystyle{\max_{1\leq j\leq
\N(\eta,\C,d_{\C})}\sup_{\varphi\in
B_{(\varphi_j,\eta)}}\!\!\!J_{\varphi,n,2}}$, observe that
\begin{eqnarray}\label{decompostion3}
\lefteqn{\max_{1\leq j\leq \N(\eta,\C,d_{\C})}\sup_{\varphi\in
B(\varphi_j,\eta)}J_{\varphi,n,2}}\nonumber\\
&\hspace*{1.5cm}\leq& \max_{1\leq j\leq
\N(\eta,\C,d_{\C})}\sup_{\varphi\in B(\varphi_j,\eta)} \max_{1\leq
k\leq d_{\epsilon, n}}|
f_{\varphi,n}(x,t^\epsilon_{\varphi_j,k})-f_{\varphi_j,n}(x,t^\epsilon_{\varphi_j,k})|\nonumber\\
&\hspace*{1.5cm}&+\max_{1\leq j\leq \N(\eta,\C,d_{\C})}
\max_{1\leq k\leq d_{\epsilon, n}}|
f_{\varphi_j,n}(x,t^\epsilon_{\varphi_j,k})-\bar{f}_{\varphi_j,n}(x,t^\epsilon_{\varphi_j,k})|
 \nonumber \\
&\hspace*{1.5cm}&+ \max_{1\leq j\leq \N(\eta,\C,d_{\C
})}\sup_{\varphi\in B(\varphi_j,\eta)} \max_{1\leq k\leq
d_{\epsilon, n}}|
\bar{f}_{\varphi_j,n}(x,t^\epsilon_{\varphi_j,k})-\bar{f}_{\varphi,n}(x,t^\epsilon_{\varphi_j,k})|\nonumber
\\&\hspace*{1.5cm}=:& J_{n,1}+J_{n,2}+J_{n,3}.
\end{eqnarray}
By the same arguments  as in the statement (\ref{J23}), we can
show, under Condition {\bf A4} (ii), that
\begin{eqnarray}\label{J24}\left(J_{n,1}+J_{n,3}\right)
\leq C_H\eta
h^{-2}\left(l_n(x)+\bar{l}_n(x)\right)=O_{a.s.}\left(\eta
h^{-2}\right).
\end{eqnarray}
We have to deal now with the middle term $J_{n,2}$. Observe that
\begin{eqnarray*}
\mathbb{P}\left(J_{n,2}>\lambda\right) &= &
\mathbb{P}\left(\max_{1\leq j\leq \N(\eta,\C,d_{\C})} \max_{1\leq
k\leq d_{\epsilon, n}}|
f_{\varphi_j,n}(x,t^\epsilon_{\varphi_j,k})-\bar{f}_{\varphi_j,n}(x,t^\epsilon_{\varphi_j,k})|>\lambda\right)\nonumber \\
&\leq & \sum_{j=1}^{\N(\eta,\C,d_{\C})} \sum_{k=1}^{d_{\epsilon,
n}}\mathbb{P}\left(\frac{1}{nh}\left|\sum_{i=1}^nL_{i,\varphi_j}(x,
t^\epsilon_{\varphi_j,k})\right| \geq \lambda\right),
\end{eqnarray*}
where $L_{i,\varphi_j}(x,y)=\frac{1}{\E\lbrack
\Delta_1(x)\rbrack}\left\lbrack\Delta_i(x)H^(\frac{y-\varphi_j(Z_i)}{h})-\mathbb{E}\left[
\Delta_i(x)H(\frac{y-\varphi_j(Z_i)}{h}) |\ {\cal
F}_{i-1}\right]\right\rbrack$. Notice that $L_{i,\varphi_j}(x,y)$
 is  a martingale difference bounded by the quantity
$\displaystyle{M:=\frac{2\bar
K\overline{H}}{\phi(h)[M_1f_1(x)+o(1)]}}.$ In fact, since the
kernel $K$ and  the function $H$ are bounded, it follows easily in
view of Lemma 2 (ii) in La\"ib \& Louani (2011) that
\begin{eqnarray*}
 |L_{i,\varphi_j}(x, y)|\leq\frac{2\bar
K\overline{H}}{\E\lbrack \Delta_1(x)\rbrack}=\frac{2\bar
K\overline{H}}{\phi(h)[M_1f_1(x)+o(1)]},
\end{eqnarray*}
where ${\overline{H}}:=\sup_{y\in\R}H(y)$ and
${\overline{K}}:=\sup_{y\in\R}K(y)$. Observe now that
\begin{eqnarray*}
 \E\lbrack (L_{i,\varphi_j}(x,t^\epsilon_{\varphi_j,k}))^{2}|\F_{i-1}\rbrack\!\leq\!\frac{1}{(\E\lbrack
\Delta_1(x)\rbrack)^2}\E\!\left\lbrack\!
\left(\!\Delta_i(x)H\left(\frac{t^\epsilon_{\varphi_j,k}\!-\!\varphi_j(Z_i)}{h}\!\right)\!\right)^{2}
\mid \F_{i-1}\right\rbrack.
\end{eqnarray*}
Therefore, by condition {\bf{A5}}, we have
\begin{eqnarray*}
\lefteqn{\E\left\lbrack
\left(\Delta_i(x)H\left(\frac{t^\epsilon_{\varphi_j,k}-\varphi_j(Z_i)}{h}\right)\right)^{2}
\mid \F_{i-1}\right\rbrack}\\
&\hspace*{2.5cm}=&\E\left\lbrack\left(\Delta_i(x)\right)^2
\E\left\lbrack\left(H\left(\frac{t^\epsilon_{\varphi_j,k}-\varphi_j(Z_i)}{h}\right)\right)^{2}|\G_{i-1}\right\rbrack
\mid
\F_{i-1}\right\rbrack\\
&\hspace*{2.5cm}=&\E\left\lbrack\left(\Delta_i(x)\right)^2
\E\left\lbrack\left(H\left(\frac{t^\epsilon_{\varphi_j,k}-\varphi_j(Z_i)}{h}\right)\right)^{2}|X_i\right\rbrack
\mid \F_{i-1}\right\rbrack\\
&\hspace*{2.5cm}=&\E\left\lbrack \left(\Delta_i(x)\right)^2
\int_{\R}\left(H\left(\frac{u}{h}\right)\right)^2
  g_{\varphi_j}(t^\epsilon_{\varphi_j,k}-u|X_i)du\mid
\F_{i-1}\right\rbrack\\
&\hspace*{2.5cm}:=&\E\left\lbrack \left(\Delta_i(x)\right)^2
\T_{i,1}\mid \F_{i-1}\right\rbrack+\T_2\E\left\lbrack
\left(\Delta_i(x)\right)^2 \mid \F_{i-1}\right\rbrack,
\end{eqnarray*}
where we have set ${\cal
T}_{i,1}=\int_{\R}\left(H\left(\frac{u}{h}\right)\right)^2
  \left(g_{\varphi_j}(t^\epsilon_{\varphi_j,k}-u|X_i)-g_{\varphi_j}(t^\epsilon_{\varphi_j,k}|x)\right)du$
  \ and \ %
  ${\cal T}_{2}=\int_{\R}\left(H\left(\frac{u}{h}\right)\right)^2
 g_{\varphi_j}(t^\epsilon_{\varphi_j,k}|x)du.
  $
Subsequently, for  $\eta>0$, we have
 \begin{eqnarray*}
\T_{i,1}&\leq&
\int_{|u|\leq\eta}\left(H\left(\frac{u}{h}\right)\right)^2
  \left |g_{\varphi_j}(t^\epsilon_{\varphi_j,k}-u|X_i)-g_{\varphi_j}(t^\epsilon_{\varphi_j,k}|x)\right|du
\\&&+  \int_{|u|>\eta}\left(H\left(\frac{u}{h}\right)\right)^2
  \left|g_{\varphi_j}(t^\epsilon_{\varphi_j,k}-u|X_i)-g_{\varphi_j}
  (t^\epsilon_{\varphi_j,k}|x)\right|du
\\&\leq&h\sup_{|u|\leq\eta}|g_{\varphi}(t^\epsilon_{\varphi_j,k}-u|X_i)-g_{\varphi_j}(
t^\epsilon_{\varphi_j,k}|x)|\int_{|u|\leq\eta/\!h}\left(H(u)\right)^2
du
\\&&+h
\sup_{|u|>\eta/\!h}(H(u))^2+hg_{\varphi_j}(t^\epsilon_{\varphi_j,k}|x)\int_{|u|>\eta/\!h}\left(H(u)\right)^2du;
\end{eqnarray*}
Condition {\bf{A3}} (iv) allows us, for any $\eta>0$, to write
\begin{eqnarray*} \T_{i,1}&\leq& hC_x(|\eta|^{\nu}+ d(x, \
X_i)^{\beta})\int_{|u|\leq\eta/\!h}\left(H(u)\right)^2 du
\\&&+h
\sup_{|u|>\eta/\!h}(H(u))^2+hg_{\varphi_j}(t^\epsilon_{\varphi_j,k}|x)\int_{|u|>\eta/\!h}\left(H(u)\right)^2du.
\end{eqnarray*}
Thus,
\begin{eqnarray*}
\E\left\lbrack \left(\Delta_i(x)\right)^2 \T_{i,1}\mid
\F_{i-1}\right\rbrack&\leq& h\E\bigg\lbrack
\left(\Delta_i(x)\right)^2\bigg(C_x(|\eta|^{\nu}+ d(x, \
X_i)^{\beta})\int_{|u|\leq\frac{\eta}{h}}H^2(u)du
\\&&+
\sup_{|u|>\frac{\eta}{h}}H^2(u)+g_{\varphi_j}(t^\epsilon_{\varphi_j,k}|x)\int_{|u|>\frac{\eta}{h}}
H^2(u)du\bigg)\mid\F_{i-1}\bigg\rbrack\\
&\leq& h\bigg(C_x(|\eta|^{\nu}+
h^{\beta})\int_{|u|\leq\eta/\!h}\left(H(u)\right)^2 du+
\sup_{|u|>\frac{\eta}{h}}H^2(u)
\\&&+g_{\varphi_j}(t^\epsilon_{\varphi_j,k}|x)\int_{|u|>\frac{\eta}{h}}H^2(u)du
\bigg)\E\bigg\lbrack
\left(\Delta_i(x)\right)^2\mid\F_{i-1}\bigg\rbrack.
\end{eqnarray*}
On another hand, we can see easily, for some positive constant
$C_0$, that
$$\T_2\E\left\lbrack
\left(\Delta_i(x)\right)^2 \mid \F_{i-1}\right\rbrack\leq
C_0h\E\left\lbrack \left(\Delta_i(x)\right)^2 \mid
\F_{i-1}\right\rbrack.$$
Therefore, since $H$ is bounded and $h\to 0$, it follows then that
there exists a constant $C_1>0$ such that
\begin{eqnarray*}
 \E\!\left\lbrack
\left(\!\Delta_i(x)H\!\left(\frac{t^\epsilon_{\varphi_j,k}-\varphi(Z_i)}{h}\right)\right)^{2}
\mid \F_{i-1}\right\rbrack &\!\leq\!& C_1 h \E\!\bigg\lbrack
\left(\!\Delta_i(x)\right)^2\mid\F_{i-1}\bigg\rbrack \ \mbox{as}\
n\rightarrow\infty.\end{eqnarray*}
Furthermore, using Condition  {\bf (A1)}, which supposes almost
surely that $f_{i,1}$ is bounded by a deterministic function
$b(x)$ and that $\psi_{i,x}(h)\leq \phi(h)$ as $h\to 0$, together
with Lemma 2 in La\"ib \& Louani (2011), we have for $n$ large enough
\begin{eqnarray*}
 \E\lbrack (L_{i,\varphi_j}(x,t^\epsilon_{\varphi_j,k}))^{2}|\F_{i-1}\rbrack &\leq& \frac{C_1h}{\left(\E\lbrack
\Delta_1(x)\rbrack\right)^{2}}\E\left\lbrack
\left(\Delta_i(x)\right)^{2} \mid \F_{i-1}\right\rbrack\nonumber \\
&\leq & \frac{C_1
h}{\phi(h)[M_1^2f_1^2(x)+o(1)]}[M_2b_i(x)+1]=:d_i^2.
\end{eqnarray*}
%
%
Moreover, using Conditions {\bf A1} (iii),(v), one may write
\begin{eqnarray*}
\frac{4D_n}{n}+2Mh\lambda &=& \frac{h}{\phi(h)}\left[
\frac{4C_1[M_2D(x)+1]}{M_1^2f_1^2(x)+o(1)}+
\frac{4\lambda\bar{K}\bar{H}}{M_1f_1(x)+o(1)}   \right]
\end{eqnarray*}
and
$\displaystyle{\frac{nh^2\lambda^2}{4D_n/n+2Mh\lambda} =
nh\phi(h)\lambda^2C_{\epsilon}(x)}$, \ where
$$
C_{\epsilon}(x)=\frac{M_1f_1(x)}{4}. \frac{1}{
\frac{C_1(M_2D(x)+1)}{M_1f_1(x)} +\lambda \bar{H}\bar{K}+o(1)}.$$
%
Consequently,

\vskip 1cm
%
$$\mathbb{P}\left(\frac{1}{nh}\left|\sum_{i=1}^nL_{i,\varphi_j}(x,
t_{\varphi_j,k}^{\epsilon})\right| \geq
\lambda\right)\leq 2\exp\left\{-nh\phi(h)\lambda^2
C_{\epsilon}(x)\right\}.$$
Choosing $\lambda=\lambda_n$ and $m_n=\eta$, we obtain
%
\begin{eqnarray*}
\mathbb{P}\left(J_{2,n} \geq\lambda_n\right)&\leq&
2\N(\eta,\C,d_{\C})d_{\epsilon,n}\exp\left\{-nh\phi(h)\lambda^2_n
C_{\epsilon}(x)\right\}\\
&=&2\exp\left\{-\lambda^2_nnh\phi(h)\left[C_\epsilon(x)-\frac{c_\epsilon\log
\N(\eta,\C,d_{\C})}{\eta\lambda_n^2nh\phi(h)}\right]\right\}.
\end{eqnarray*}
Taking into account the condition (\ref{vc}), it suffices to use
the Borel-Cantelli Lemma to conclude the proof.
%
%
\hfill$\Box$ \vspace{3mm}


\noindent{\bf Proof of Theorem \ref{th1}}.
 Taylor series expansion of the function $g_\varphi(\hat{\Theta}_{\varphi,n}(x)|x)$ around $\Theta_\varphi(x)$ together with the definition of $\Theta_\varphi(x)$ yield
\begin{eqnarray}\label{eqnt1}
g_\varphi(\hat{\Theta}_{\varphi,n}(x)|x)=g_\varphi(\Theta_\varphi(x)|x)+(\hat{\Theta}_{\varphi,n}(x)-\Theta_\varphi(x))^{2}\frac{1}{2}
g_\varphi^{(2)}(\Theta^{*}_{\varphi,n}(x)|x),
\end{eqnarray}
where $\Theta^{*}_{\varphi,n}$ is between
$\hat{\Theta}_{\varphi,n}(x)$ and $\Theta_\varphi(x)$.
Subsequently, considering the statement (\ref{eqnt1}) we obtain
\begin{eqnarray}\label{eqnt3}
(\hat{\Theta}_{\varphi,n}(x)-\Theta_\varphi(x))^{2}|g^{(2)}(\Theta^{*}_{\varphi,n}(x)|x)|=O\left(\sup_{y
\in S_\varphi}|g_n(y|x)-g(y|x)|\right).
\end{eqnarray}

To end the proof of the theorem, we need the following lemma which deals with the
uniform (with respect to $\varphi \in{\cal C}$) asymptotic behavior of the conditional mode estimate.

\begin{lemma}\label{lemmethetan}
Under assumptions of Proposition \ref{lemmasup}, we have
\begin{eqnarray*}
\lim_{n\rightarrow\infty}\sup_{\varphi\in \C}
|\hat{\Theta}_{\varphi,n}(x)-\Theta_{\varphi}(x)|=0 \quad a.s.
\end{eqnarray*}
\end{lemma}
\noindent{\it Proof of Lemma} \ref{lemmethetan}. Since by the
assumption {\bf A3}(ii), uniformly in $\varphi\in\C$,
$g_\varphi(\cdot| x)$ is uniformly continuous on the compact set
$S_\varphi$ on which $\theta_\varphi(x)$ is the unique mode.  Then,
proceeding as in Parzen (1962), for any $\varepsilon>0$, there
exists $\zeta> 0$ such that, for any $y\in S_\varphi$,
\begin{eqnarray}\label{condition_Mode}
\sup_{\varphi\in\C}|\Theta_{\varphi}(x)-y|\geq\epsilon \Rightarrow
\sup_{\varphi\in\C}|g_{\varphi}(\Theta_{\varphi}(x)|x)-
g_{\varphi}(y|x)|\geq \zeta.
\end{eqnarray}
On another hand, we have
\begin{eqnarray}\label{sup1}
\sup_{\varphi\in\C}|g_\varphi(\hat{\Theta}_{\varphi,n}(x)|x)-g_\varphi(\Theta_\varphi(x)|x)|
&\leq&
\sup_{\varphi\in\C}|g_{\varphi,n}(\hat{\Theta}_{\varphi,n}(x)|x)-g_\varphi(\hat{\Theta}_{\varphi,n}(x)|x)|
\nonumber \\
&+&
\sup_{\varphi\in\C}|g_{\varphi,n}(\hat{\Theta}_{\varphi,n}(x)|x)-g_\varphi(\Theta_{\varphi}(x)|x)|
\nonumber \\
&\leq&
\sup_{\varphi\in\C}\sup_{y\in
S_\varphi}|g_{\varphi,n}(y|x)-g_\varphi(y|x)|\nonumber\\
&&+ \sup_{\varphi\in\C}|\sup_{y\in
S_\varphi}g_{\varphi,n}(y|x)-\sup_{y\in S_\varphi}g_\varphi(y|x)|
\nonumber \\
&\leq &  2\sup_{\varphi\in \C}\sup_{y \in
S_\varphi}|g_{\varphi,n}(y|x)-g_\varphi(y|x)|.
\end{eqnarray}
Using the statements (\ref{condition_Mode}) and (\ref{sup1})
combined with Proposition \ref{lemmasup}, we obtain the result.
\hfill$\Box$ \vspace{3mm}
%
%

We come back now on the proof of the Theorem.  Making use of Lemma 7 combined
with conditions A3(iii)-(iv), we deduce that

\begin{eqnarray}\label{eqnt4}
\lim_{n  \rightarrow\infty}
\sup_{\varphi\in\C}|g^{(2)}(\Theta^{*}_{\varphi,n}(x)|x)|=\sup_{\varphi\in\C}|g^{(2)}(\Theta_{\varphi}(x)|x)|=\Phi(x)\neq
0.
\end{eqnarray}
Moreover, the statements (\ref{eqnt3}), (\ref{eqnt4}) imply that
\begin{eqnarray}\label{eqnt5}
\sup_{\varphi\in\C}(\hat{\Theta}_{\varphi,n}(x)-\Theta_\varphi(x))^{2}=O\left(\sup_{\varphi\in\C}\sup_{y
\in S_\varphi}|g_n(y|x)-g(y|x)|\right),
\end{eqnarray}
which is enough, while considering  Proposition \ref{lemmasup}, to
complete the proof of Theorem \ref{th1}. \hfill$\Box$ \vspace{3mm}


\end{document}